\documentclass[times,twocolumn,longtitle]{elsarticle}
\usepackage{jasr}
\usepackage{framed,multirow}
\usepackage[utf8]{inputenc}
\usepackage[T1]{fontenc}
\usepackage{amssymb}
\usepackage{latexsym}
\usepackage{amsmath}    %
\usepackage{bm}        
\usepackage{graphicx}  
\usepackage{pdflscape}
\usepackage{booktabs}   
\usepackage{siunitx}  
\usepackage{nicematrix}
\usepackage{caption}
\usepackage{subfig}
\usepackage[switch]{lineno}
\usepackage{url}
\usepackage{xcolor}
\usepackage{orcidlink}
\usepackage{geometry}
\definecolor{newcolor}{rgb}{.8,.349,.1}
\usepackage{hyperref}
\usepackage{comment}

\journal{Advances in Space Research}
\begin{document}

\verso{Nasser \textit{et al.}}
\begin{frontmatter}
\title{Detailed Analysis of the NGC 2168 Cluster, Leveraging Gaia DR3}
\author[1]{Nasser M. \snm{Ahmed}\corref{cor1} \,\orcidlink{0000-0002-7701-3032}}
\cortext[cor1]{Corresponding author}
\ead{nasser_ahnmed@yahoo.com}
\author[2]{Remziye \snm{Canbay}\orcidlink{0000-0003-2575-9892}}
\ead{rmzycnby@gmail.com}
\author[3]{Deniz Cennet \snm{{\c{C}}\i nar}\orcidlink{0000-0001-7940-3731}}
\ead{denizcennetcinar@gmail.com}
\affiliation[1]{organization={National Research Institute of Astronomy and Geophysics (NRIAG)},
                addressline={11421 Helwan},
                city={Cairo},
                country={Egypt}}

\affiliation[2]{organization={Department of Astronomy and Space Sciences, Faculty of Science, Istanbul University},
                postcode={34119},
                city={Istanbul},
                country={Turkey}}
\affiliation[3]{organization={Programme of Astronomy and Space Sciences, Institute of Graduate Studies in Science, Istanbul University},
                postcode={34116},
                city={Istanbul},
                country={Turkey}}
\received{\today}
\finalform{}
\accepted{}
\availableonline{}
\communicated{}

\begin{abstract}
NGC~2168 (M35) serves as a fundamental benchmark for studying stellar evolution and dynamical environments at the transition between young and intermediate-age populations. We present a comprehensive analysis of the cluster's kinematic, structural, and astrophysical properties utilizing high-precision astrometry and photometry data from \textit{Gaia} Data Release~3 (DR3), complemented by 2MASS data. A statistical membership assessment yields a clean sample of probable members (N $\approx$ 1397), with mean proper motion components of $\mu_{\alpha}\cos\delta = 2.278 \pm 0.006~\mathrm{mas\,yr^{-1}}$ and $\mu_{\delta} = -2.893 \pm 0.006~\mathrm{mas\,yr^{-1}}$, along with a mean trigonometric parallax of $\varpi = 1.154 \pm 0.052~\mathrm{mas}$. We derived the cluster's fundamental parameters via isochrone fitting, determining an age of $190 \pm 12~\mathrm{Myr}$, a metallicity of $\mathrm{[M/H]} \approx -0.048~\mathrm{dex}$, and a probabilistic distance of $840 \pm 54~\mathrm{pc}$. The radial density profile is well described by a generalized King model with $\beta=1$ ($r_{c} = 7.97'$, $r_{\mathrm{cl}} = 36.69'$), revealing the presence of a loosely bound, extended stellar halo. Furthermore, we detect a spatial elongation oriented perpendicular to the Galactic plane, likely a signature of vertical tidal heating or disk shocking. The mass function analysis exhibits a multimodal Gaussian structure, suggesting a complex dynamical formation history beyond a simple power-law distribution. Finally, orbital integration confirms NGC~2168 as a thin disk object with a maximum vertical excursion of $\sim 171~\mathrm{pc}$, consistent with the observed vertical morphological deformation.
\end{abstract}
\end{frontmatter}

\section{Introduction}
Open clusters (OCs) have long served as fundamental benchmarks for constraining theories of stellar evolution and dynamics. As gravitationally bound systems formed from the same molecular cloud, member stars share a common distance, age, and initial chemical composition. This homogeneity allows OCs to act as excellent tracers of the Galactic disk's structure and chemical evolution \citep{Friel1995, Dias2002}. Furthermore, rich clusters provide unique laboratories for investigating dynamical processes such as mass segregation, evaporation, and the impact of the Galactic potential on star clusters over time.

The advent of the $Gaia$ mission has fundamentally transformed the study of these systems. The third data release ($Gaia$ DR3) offers unprecedented astrometric precision, extending to faint magnitudes and providing comprehensive radial velocity measurements for a significant subset of stars \citep{GaiaCollaboration2023}. Unlike earlier photometric surveys limited by field star contamination, $Gaia$’s high-precision proper motions and parallaxes allow for a robust separation of cluster members from the field population in high-dimensional phase space. This capability is critical for analyzing the internal kinematics and identifying extended stellar halos or tidal tails, which are often masked by background noise in purely photometric studies \citep{Cantat-Gaudin2020, Tarricq2021}

NGC 2168 (M35), located in the constellation Gemini, represents one of the most populous and well-studied intermediate-age open clusters in the solar neighborhood. With an age of approximately 100–150 Myr and a distance of $\approx$800–900 pc, it serves as a critical template for studying stellar properties at the transition between young and intermediate-age populations \citep{Sung1999, Kalirai2003}. Previous investigations have extensively characterized its photometric properties and binary fraction \citep{Geller2010, Leiner2015}. While recent large-scale surveys utilizing \textit{Gaia} data \citep{Cantat-Gaudin2020, hunt2024improving} have provided updated fundamental parameters, these works primarily focused on cataloging mean properties across many clusters rather than conducting a deep structural analysis of individual targets. Consequently, the cluster's extended morphological features, such as its distant halo and complex mass function substructures, remain largely unexplored. In this study, we conduct a dedicated analysis of NGC 2168 leveraging the homogeneity and depth of \textit{Gaia} DR3 to resolve these detailed structural and dynamical properties.

\begin{table*}[t]
    \centering
    \caption{Comparison of the fundamental parameters of NGC 2168 derived in this study with selected literature values. The table is sorted chronologically.}
    \label{tab:literature}
    \setlength{\tabcolsep}{5pt} 
    \begin{tabular}{lcccccc}
        \toprule
        \toprule
        Reference & Age & Distance & $E(B-V)$ & $\mu_{\alpha}\cos\delta$ & $\mu_{\delta}$ & $N$ \\
         & (Myr) & (pc) & (mag) & (mas yr$^{-1}$) & (mas yr$^{-1}$) & (stars) \\
        \midrule
        \citet{Sung1999} & $\sim200$ & $830$ & $0.255$ & -- & -- & -- \\
        \citet{Dias2002} & $\sim178$ & $912$ & $0.200$ & $0.65$ & $-3.06$ & 813 \\
        \citet{Kalirai2003} & $\sim178$ & $912$ & $0.200$ & -- & -- & -- \\
        \citet{Kharchenko2013} & $135$ & $805$ & $0.171$ & $2.22$ & $-2.94$ & 1047 \\
        \citet{Bouy2015} & $150$ & $824$ & $0.260$ & $2.45$ & $-2.78$ & 4349 \\
        \citet{Cantat-Gaudin2020} & $178$ & $884$ & $0.240$ & $2.28$ & $-2.95$ & 1445 \\
        \citet{hunt2024improving} & $166$ & $837$ & $0.191$ & $2.29$ & $-2.96$ & 1609 \\
        \midrule
        \textbf{This Study} & $\mathbf{190 \pm 12}$ & $\mathbf{840 \pm 54}$ & $\mathbf{0.31 \pm 0.04}$ & $\mathbf{2.28 \pm 0.24}$ & $\mathbf{-2.89 \pm 0.23}$ & \textbf{1397} \\
        \bottomrule
    \end{tabular}
\end{table*}

In this study,  we revisit NGC 2168 leveraging the homogeneity and depth of $Gaia$ DR3. Our primary objective is to refine the cluster’s fundamental parameters and investigate its dynamical evolution with a strictly defined membership list. By combining precise astrometry with available photometric data, we derive the cluster’s mass function and examine evidence for mass segregation. We also update the cluster's distance and age estimates through isochrone fitting to the cleaned color-magnitude diagram (CMD).

The paper is structured as follows: Section 2 describes the $Gaia$ DR3 and 2MASS data, as well as the adopted treatment of field contamination and the comparison field selection. Section 3 presents the cluster structure analysis based on the radial density profile and the corresponding King-model fit. Section 4 details the membership determination, including the HDBSCAN-based selection and the subsequent probability-based refinement. Section 5 derives the photometric properties of NGC 2168, including the extinction analysis and the isochrone-based estimates of distance, age, and metallicity. Section 6 investigates the cluster mass and the present-day mass function, including the multi-component representation of the mass distribution. Section 7 discusses the kinematics and dynamics of the cluster and its Galactic orbit. Finally, Section 8 summarizes the main results and places NGC 2168 in the broader context of open-cluster evolution.

\section{Data} \label{sec:data}
\textcolor{black}{In this study, we utilize two comprehensive and complementary datasets, $Gaia$ DR3 and 2MASS, to conduct a detailed analysis of the open cluster NGC~2168. The combination of high-precision astrometry from $Gaia$ and near-infrared photometry from 2MASS provides a robust foundation for isolating cluster members, deriving accurate astrophysical parameters, and investigating the cluster's structural and kinematic properties.}

\begin{figure}[t]
\centering
\includegraphics[width=0.95\linewidth]{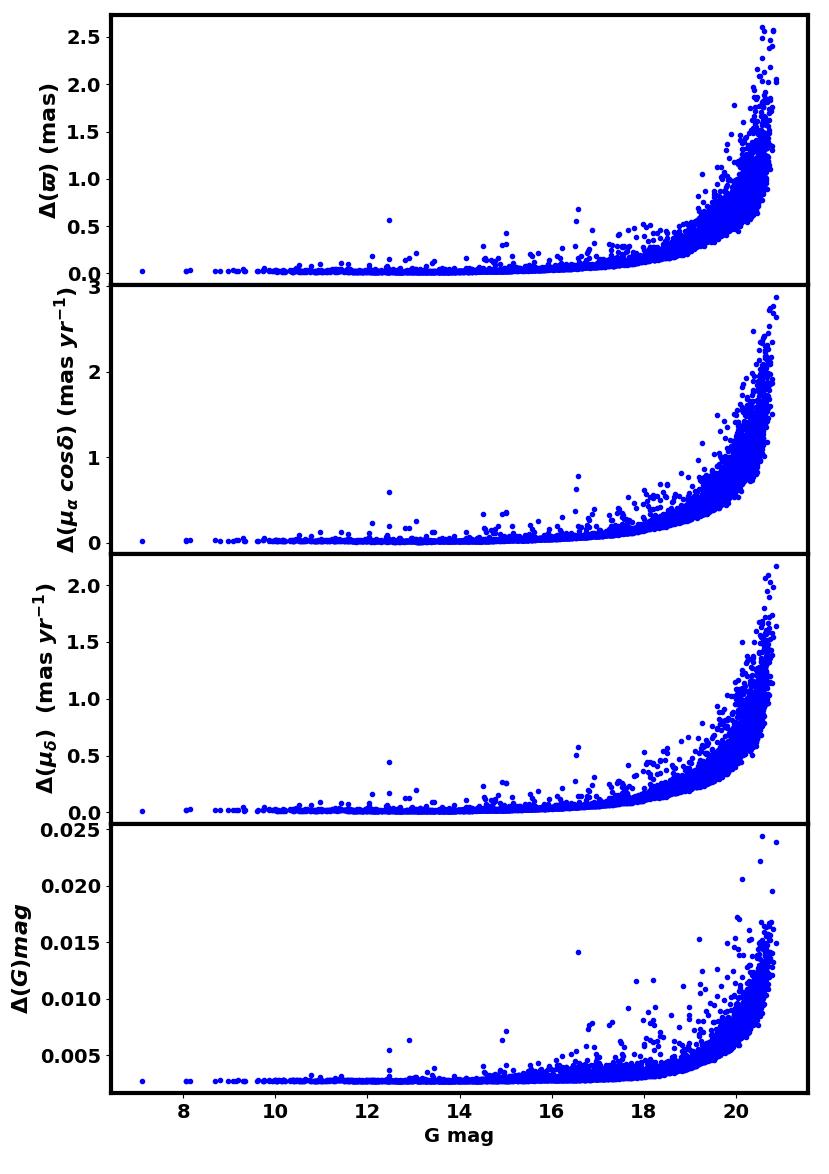}
\caption{Uncertainties in the astrometric parameters (trigonometric parallax and proper-motion components) and in the photometric $G$ magnitude as a function of $G$ magnitude for stars in the field of NGC~2168.}
   \label{fig:error}
\end{figure}

\textcolor{black}{Astrometric and photometric data for NGC~2168 were retrieved from the $Gaia$ Data Release 3 (DR3) catalog \citep{GaiaCollaboration2023}. The dataset includes five-parameter astrometry, comprising sky positions ($\alpha$, $\delta$), proper motion components ($\mu_{\alpha}\cos\delta$, $\mu_{\delta}$), and trigonometric parallaxes ($\varpi$), along with precise photometry, adopting a limiting magnitude of $G = 21$~mag. $Gaia$ DR3 provides crucial astrophysical parameters derived from these measurements, including broad-band photometry and mean radial velocity spectra for bright sources. The astrometric precision depends on the magnitude of the sources: for bright stars ($G \leq 17$~mag), parallax uncertainties are 0.02–0.07~mas, increasing to $\sim$0.5~mas at $G = 20$~mag and 1.3~mas at $G = 21$~mag. Proper motion errors follow a similar trend, from 0.02–0.07~mas~yr$^{-1}$ for bright stars up to 1.4~mas~yr$^{-1}$ at the faint end. The distribution of these uncertainties is illustrated in Fig.~\ref{fig:error}.}

\textcolor{black}{Near-infrared photometry from the Two Micron All-Sky Survey (2MASS; \cite{Skrutskie2006}) was also used to refine the interstellar extinction and cluster age. This catalog provides uniform $JHK_s$ measurements for a large number of stars across the sky, which are particularly useful for studies of open clusters such as NGC~2168.}

\subsection{Contamination and Field Selection} \label{sec:data_lim}

\begin{figure*}[t]
\centering
\subfloat[$0 < \varpi < 1.4$~mas]{%
\label{fig:surf1}
\includegraphics[width=0.32\linewidth]{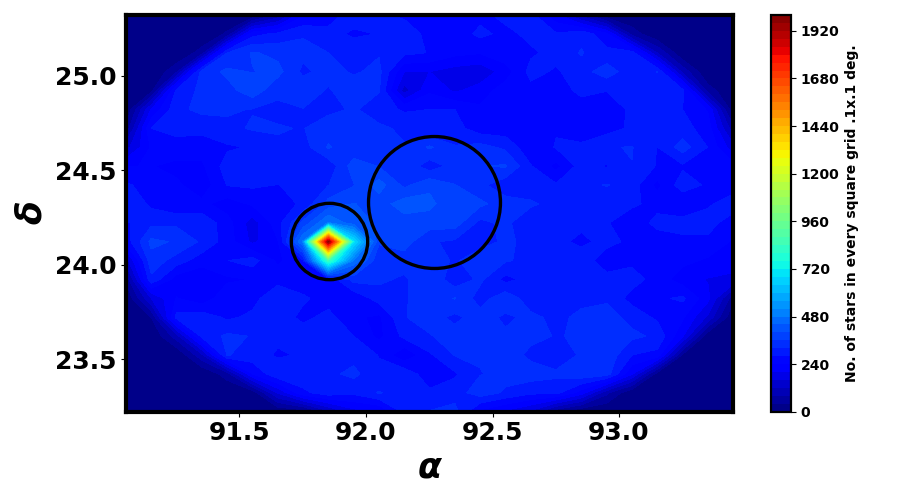}
}\hfill
\subfloat[$0.9 < \varpi < 1.4$~mas]{%
\label{fig:surf2}
\includegraphics[width=0.32\linewidth]{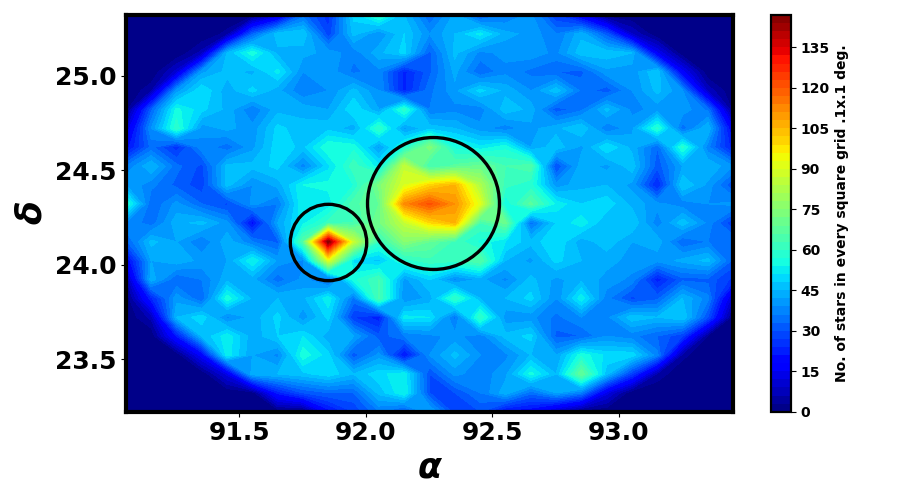}
}\hfill
\subfloat[$0.9 < \varpi < 1.4$~mas,\ $1.4 < \mu_{\delta} < 3.2$~mas~yr$^{-1}$]{%
\label{fig:surf3}
\includegraphics[width=0.32\linewidth]{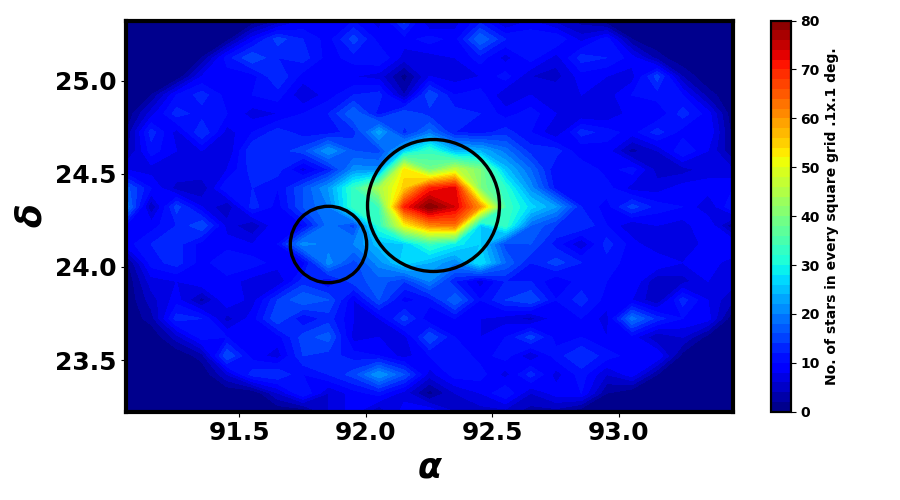}
}

\caption{Surface number density maps in the field of NGC~2168 under different astrometric selection criteria. In the first panel, two stellar overdensities are visible, corresponding to NGC~2168 (larger circle) and the more distant background cluster NGC~2158 (smaller circle).  Applying a stricter parallax cut suppresses the contribution from NGC~2158, while the combined parallax and proper-motion selection effectively removes the background cluster, yielding a clean density distribution dominated by NGC~2168.}
\label{Fig:surf}
\end{figure*}

Our analysis of the NGC~2168 field is significantly complicated by the presence of NGC~2158, a rich, intermediate-age OC ($\tau \sim 1.95$\,Gyr) located in the background at a distance of approximately 3733\,pc \citep{Nasser2025c}. With a projected separation of only $\sim$25\,arcmin from the center of NGC~2168 and a high central stellar density of $\sim$148\,stars\,arcmin$^{-2}$, NGC~2158 introduces substantial spatial contamination (see Fig.~\ref{fig:surf1}). Since spatial filtering alone is insufficient to disentangle these two populations, we utilized the distinct astrometric signatures of the clusters. 

Given the large distance contrast between the target and the background cluster, a stringent selection in parallax and proper motion space was applied. We restricted our sample to stars within the parallax range of $0.9 \leq \varpi ~\text{(mas)} \leq 1.4$ and proper motion in declination of $1.4 \le |\mu_{\delta}| ~\text{(mas\,yr$^{-1}$)} \le 3.2$. As demonstrated in Fig.~\ref{fig:surf3}, applying these astrometric cuts effectively eliminates the background contamination from NGC~2158, yielding a clean sample of NGC~2168 members for subsequent analysis.

\section{Cluster Structure and Radial Density Profile} 
\label{sec:dpr}

The structural characterization of NGC~2168 requires an accurate determination of its center and a well-constrained radial density profile (RDP). We first identified the cluster center by constructing a two-dimensional density map in right ascension and declination using \textit{Gaia}~DR3 sources. A binned density grid produced with the \texttt{numpy} \texttt{histogram2d} routine was used to locate the region of maximum stellar concentration. Repeating this procedure for the high-probability members (Section~\ref{sec:prom}) yielded consistent central coordinates, ensuring that the adopted center is not biased by field contamination. \textcolor{black}{The projected stellar density was computed in concentric annuli around the adopted center. For each ring, the density is
\begin{equation}
    n_i = \frac{N_i}{A_i},
\end{equation}
where $N_i$ is the number of stars in the $i$-th annulus and 
$A_i = \pi\,(R_{i+1}^2 - R_i^2)$ is its area \citep{Kholopov1969}.}

\textcolor{black}{Instead of adopting the simple arithmetic mean radius, the representative radius of each annulus was computed using the area-weighted expression
\begin{equation}
    r_i = \frac{2}{3}\,\frac{R_{i+1}^3 - R_i^3}{R_{i+1}^2 - R_i^2},
\end{equation}
which corresponds to the mean radius for a linear variation of surface density within the interval. This approach reduces the so-called "interval error" that may arise in regions with strong surface-density gradients.}

\textcolor{black}{The total density profile can be expressed as
\begin{equation}
    n_t(r) = n_{bg} + n_c(r),
\end{equation}
where $n_{bg}$ is the background field density and $n_c(r)$ is the intrinsic cluster component. The empirical surface-density profile of \citet{King1966} is given by
\begin{equation}
    n_c(r) = \frac{n_0}{1+(r/r_c)^2},
\end{equation}
where $n_0$ and $r_c$ denote the central density and core radius, respectively. The best-fitting parameters were determined through a $\chi^2$ minimization procedure. Although this model provides a reasonable description for many systems, its lack of an explicit truncation term may limit its applicability in the outer regions of extended OCs.}

To obtain a more realistic representation of the outer regions, we adopt a generalized form of the \cite{King1962} profile
\begin{equation}
    n_t(r)=
    \begin{cases}
        n_{bg} + k\left[ 
        \dfrac{1}{\sqrt{1+(r/r_c)^2}}
        - 
        \dfrac{1}{\sqrt{1+(r_{cl}/r_c)^2}}
        \right]^{\beta}, & r \le r_{cl},\\[8pt]
        n_{bg}, & r > r_{cl},
    \end{cases}
    \label{eq:king62}
\end{equation}
where $r_{cl}$ is the radius at which the cluster density reaches zero, and
\begin{equation}
    k = n_0 
    \left[
    1 - 
    \frac{1}{\sqrt{1+(r_{cl}/r_c)^2}}
    \right]^{-\beta}.
\end{equation}

\begin{figure*}
\centering
\includegraphics[width=0.70\linewidth]{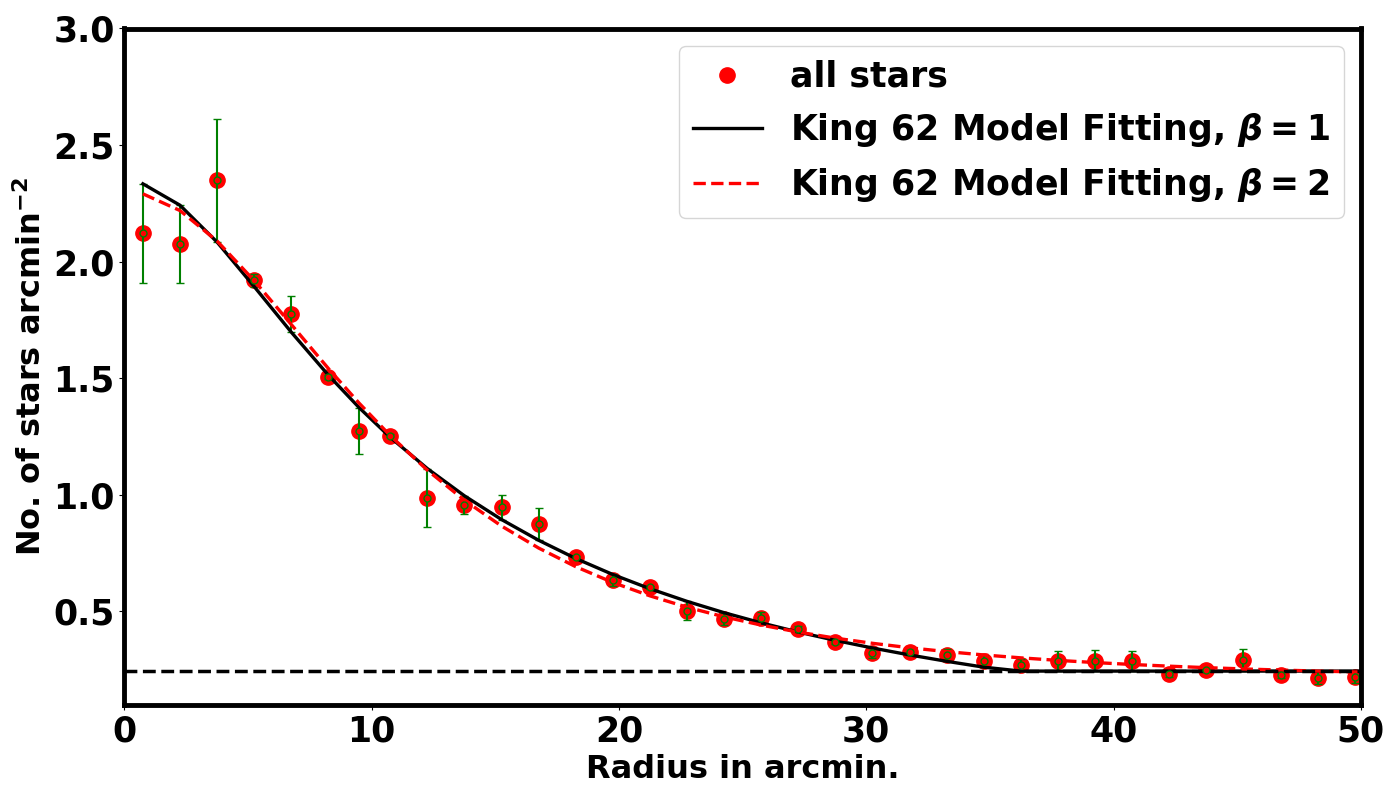}
\caption{Radial density profile of NGC~2168. The solid black curve represents the best-fit generalized King profile with $\beta = 1$, while the red dashed curve illustrates the $\beta = 2$ solution. The horizontal dashed line marks the background density level. Notably, the statistical preference for the $\beta=1$ model indicates a shallower density fall-off in the outer regions, revealing the presence of an extended stellar halo that reaches beyond the classical tidal radius estimates.}
\label{fig:rdp}
\end{figure*}

\textcolor{black}{The exponent $\beta$ determines the steepness of the cluster’s outer profile. We tested both $\beta = 1$ and $\beta = 2$ models, and the selection of $\beta = 1$ is supported by statistical criteria: it produces lower residuals and a reduced chi-square, indicating a better fit to the observed radial density profile (Fig. \ref{fig:rdp}). While the cluster radius derived in our analysis, $r_{\rm cl} = 45$ arcmin, is in close agreement with \citet{hunt2024improving}, the core radius shows a noticeable difference. This discrepancy arises primarily from the modeling approach: \citet{hunt2024improving} adopted a fixed $\beta = 2$ profile, whereas our generalized model allows $\beta$ to vary, accommodating extended or shallow halos. Consequently, the variable-$\beta$ fit yields a somewhat smaller $r_{\rm c}$, reflecting the flexibility required to represent the cluster’s structural properties accurately. The number of members in each annular bin is then calculated as}
\begin{equation}
    N_{cl,i} = n_c(r_i)\,A_i,
\end{equation}
and the total cluster population within $r_{cl}$ is
\begin{equation}
    N_{cl} = 
    \sum_{r \le r_{cl}} 
    \left(N_i - n_{bg}\,A_i\right).
\end{equation}
This provides a profile-based estimate of the cluster size that is independent of any probability threshold applied in the membership analysis.

Fitting Equation~\ref{eq:king62} to the observed stellar surface density distribution yields the structural parameters listed in Table~\ref{tab:king}. The best-fitting background density is found to be $n_{\rm bg} = 0.29 \pm 0.03$ stars arcmin$^{-2}$. The central density and core radius are $n_0 = 2.16 \pm 0.22$ stars arcmin$^{-2}$ and $r_c = 7.97 \pm 1.20$ arcmin, respectively.

The derived cluster radius, $r_{\rm cl} = 36.69 \pm 1.86$ arcmin, defines the outer boundary at which the stellar density becomes indistinguishable from the surrounding field, indicating that the cluster contribution is negligible beyond this radius. Integrating the best-fitting density profile within $r_{\rm cl}$ yields a total of $N_s$ likely cluster members.

From the fitted parameters, we further derived the concentration parameter, $C = r_{\rm cl}/r_c = 4.60$, and the density contrast parameter, $\delta_c = 1 + n_0/n_{\rm bg} = 8.45$. These values indicate a moderately to highly concentrated stellar system that is well distinguished from the background field. Parameter uncertainties were estimated from the covariance matrix returned by the \texttt{scipy} \texttt{curve\_fit} optimizer, providing formal confidence intervals on the fitted quantities.

\begin{table*}[t]
    \centering
    \caption{Best-fit structural parameters of NGC~2168 derived from the generalized King profile.
    \label{tab:king}}
    \setlength\tabcolsep{6pt} 
    \begin{tabular}{ccccccc}
        \hline \hline
        $n_0$ & $n_{\rm bg}$ & $r_c$ & $r_{\rm cl}$ & $N_{cl}$ & $C$ & $\delta_c$ \\
        (stars arcmin$^{-2}$) & (stars arcmin$^{-2}$) & (arcmin) & (arcmin)&   &  &  \\
        \hline
        $2.16 \pm 0.22$ & $0.29 \pm 0.03$ & $7.97 \pm 1.20$ & $36.69 \pm 1.86$& 1397 & 4.60 & 8.45 \\
        \hline
    \end{tabular}
\end{table*}


The derived concentration parameter ($C = 4.60$) and density contrast ($\delta_c = 8.45$) indicate that NGC~2168 is a moderately to highly concentrated OC with a well-defined central over-density. Such structural properties are commonly associated with clusters that have experienced significant internal dynamical evolution driven by two-body relaxation and mass segregation processes \citep{King1966}. In particular, the relatively high value of $C$ suggests efficient redistribution of stellar orbits toward the cluster centre, while the elevated $\delta_c$ reflects the cluster’s ability to remain gravitationally bound against the surrounding Galactic field \citep{Bonatto2007}. If the cluster age exceeds its relaxation timescale, these characteristics are consistent with a dynamically relaxed system whose present-day structure has been shaped predominantly by long-term internal dynamics rather than primordial conditions alone.

When compared with structurally similar Galactic OCs, the values of $C$ and $\delta_c$ derived for NGC~2168 fall within the upper range reported for intermediate-age and dynamically evolved systems. Previous studies have shown that dynamically young or sparsely populated OCs typically exhibit low concentration parameters ($C \lesssim 3$) and modest density contrasts ($\delta_c \lesssim 5$), whereas more evolved clusters display increasingly concentrated cores and higher contrast with respect to the background stellar field \citep{Bonatto2007, Bisht2025}. 


\section{Membership Determination} \label{sec:prom}
The assessment of fundamental parameters for star clusters is commonly complicated by the presence of field star contamination.  Historically, the determination of cluster membership relied on photometric and kinematic data. 
In light of the astrometric data provided by the $Gaia$ survey, the reliability of kinematic approaches for ascertaining membership has seen substantial improvement. Proper motion and parallax data are particularly effective in distinguishing field stars from cluster members, as stars in a cluster tend to share similar kinematic properties and distances \cite{Rangwal2019}. In this study, we utilized \textit{Gaia DR3} proper motion and parallax data to identify cluster members.
%
\subsection{The Membership Method: HDBSCAN Algorithm}
In this study, we adopted the Hierarchical Density-Based Spatial Clustering of Applications with Noise (HDBSCAN) algorithm to identify overdensities in astrometric parameter space and to robustly separate cluster members from the field population. HDBSCAN is a hierarchical extension of density-based clustering designed to recover structures across multiple density levels and to label outliers as noise \citep{Campello2013,McInnes2017}. These characteristics are particularly advantageous in open-cluster fields where the contrast between the cluster and the surrounding field can vary with position and magnitude.

The HDBSCAN-based clustering step was implemented within the \texttt{pyUPMASK} framework\footnote{https://github.com/msolpera/pyUPMASK} \citep{pyupmask}, which is a refined Python realization of the Unsupervised Photometric Membership Assignment in Stellar Clusters (UPMASK) approach originally introduced by \citet{Krone-Martins2014}. UPMASK provides a non-parametric and unsupervised strategy for membership assignment, eliminating the need for an a priori selection of field stars. The \texttt{pyUPMASK} package further extends the original concept by allowing the use of multiple clustering engines from the \texttt{scikit-learn} ecosystem \citep{Pedregosa2011}\footnote{https://scikit-learn.org/stable/} (e.g., HDBSCAN, OPTICS, KMS, Gaussian Mixture Models, and Mini Batch K-means; see \citealt{pyupmask} for references), enabling a flexible analysis of unlabeled data. Although the original UPMASK formulation emphasizes photometric information, in the present work, we applied the same unsupervised, non-parametric framework to \textit{Gaia} astrometric features (positions, proper motions, and parallaxes). 

Conceptually, HDBSCAN relaxes the assumption of a single, global density threshold and explores clustering solutions over a range of density levels within a hierarchical representation \citep{Campello2013}. This makes it more suitable than classical DBSCAN in cases where the intrinsic density of the cluster and the level of field contamination vary across the dataset. The algorithm also provides an effective mechanism for identifying and excluding noise/outliers, thereby improving the robustness of the resulting membership determination. 

\begin{figure} 
   \centering
   \includegraphics[width=1\linewidth, angle=0]{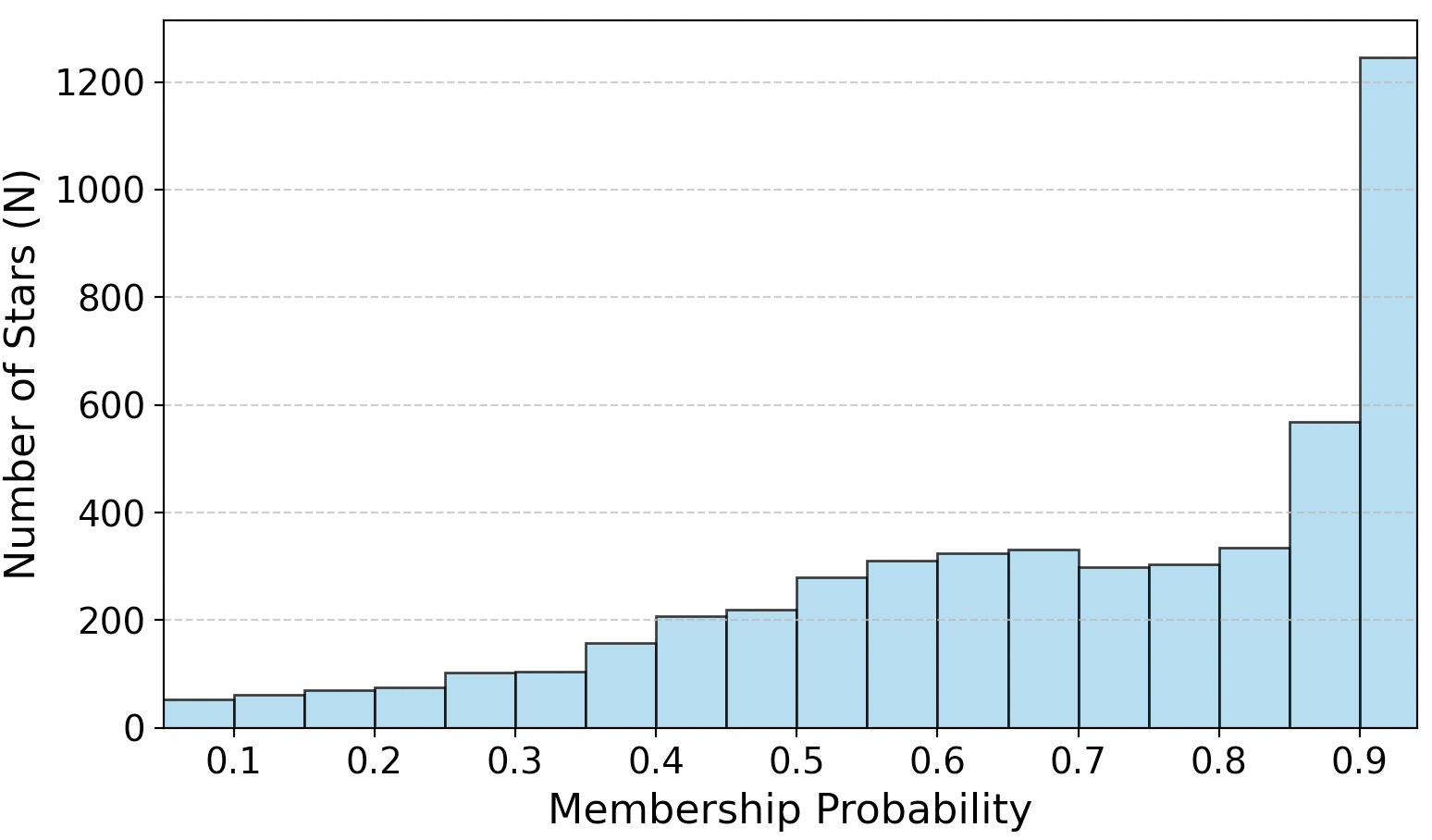}
   \caption{\textcolor{black}{Histogram of the membership probabilities for stars in the NGC 2168 region, derived from the pyUPMASK algorithm.}}
   \label{Fig:prob}
\end{figure}

We computed membership probabilities using \textit{Gaia} DR3 data ($\alpha$, $\delta$, $\mu_{\alpha}\cos\delta$, $\mu_{\delta}$, and $\varpi$) for approximately 130{,}238 sources within a radius of $50^\prime$ around NGC~2168. \textcolor{black}{Figure~\ref{Fig:prob} presents the frequency distribution of the membership probabilities ($P$) derived from the \texttt{pyUPMASK}+HDBSCAN analysis. The histogram illustrates a gradual increase in the number of stars towards higher probability values, culminating in a prominent peak for highly probable cluster members at $P \geq 0.9$.}

%
\subsection{The Probability Cut-off Value} \label{sec:cutoff}
A membership probability cut-off of 50\% is commonly used to define cluster membership; however, a single fixed threshold is not necessarily appropriate in every case. The most suitable cut-off depends on the adopted membership method and on observational conditions, particularly the density of the surrounding field and the projected distance of a star from the cluster centre. In addition, the optimal probability threshold may vary from one cluster to another. Therefore, the probability cut-off should be selected and justified carefully, because an unsuitable threshold can lead to the inclusion of contaminants and/or the rejection of genuine members. Recent studies indeed illustrate a range of adopted cut-off values. For example, \cite{Tarricq2022}, using the same HDBSCAN technique employed here, adopted a 50\% cut-off. In contrast, \cite{Zhong2022} applied UPMASK with $P>70\%$, while \cite{Gao2020} used a GMM-based selection with $P>80\%$. Consequently, the choice of an optimal probability cut-off remains a topic of ongoing discussion. \textcolor{black}{Moreover, in many previous studies, the radial density profile (RDP) constructed from stars selected using a fixed membership probability threshold fails to reproduce the expected King surface-density distribution. Since the RDP provides the fundamental structural constraints of the cluster (e.g., core radius, background level, and radial extent), this inconsistency reveals a direct mismatch between membership selection and the intrinsic structural properties of the system.}

\begin{figure*}[t]
\centering
\includegraphics[width=0.8\linewidth]{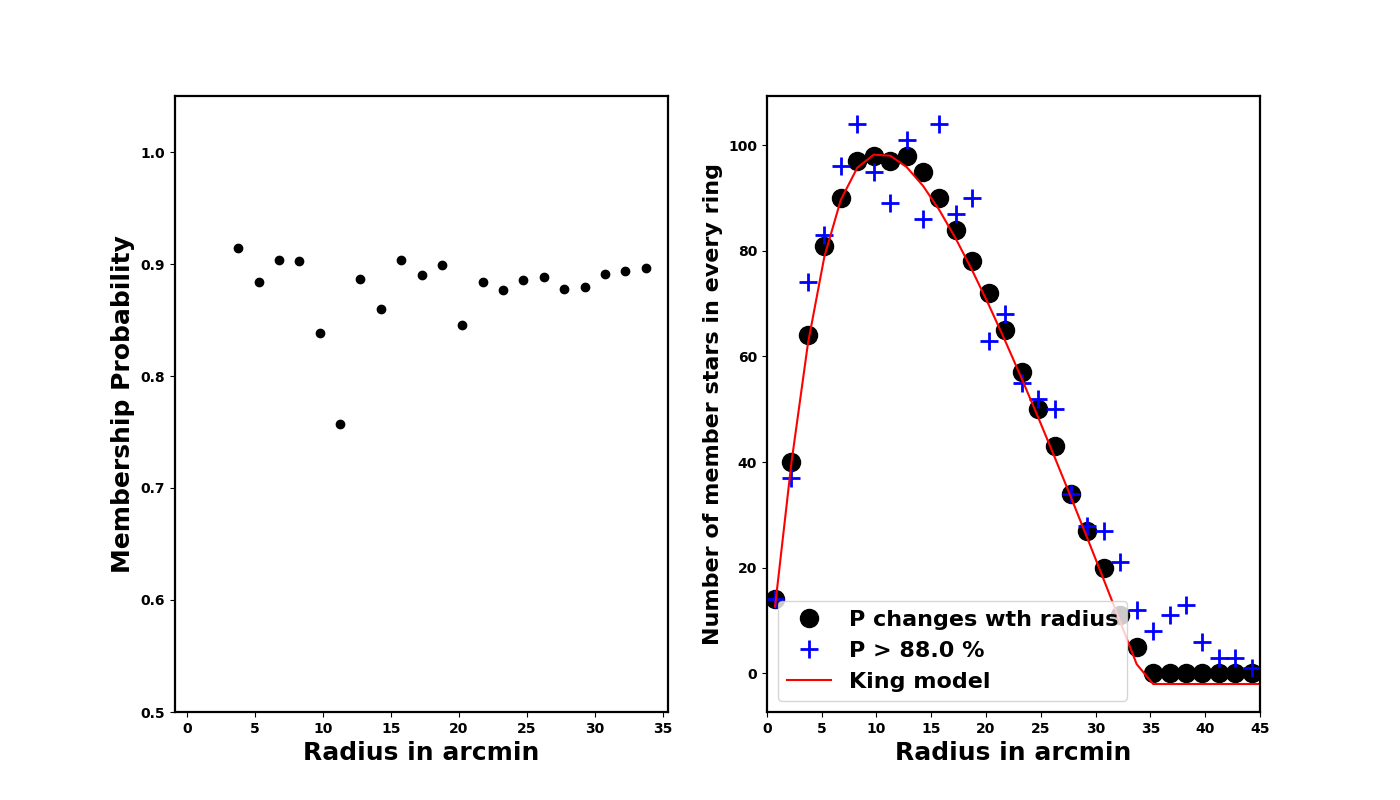}
\caption{Our methodology involves analyzing the Probability $P_{i}$ at every ring as a function of the radius $r_i$.} 
\label{fig:Prob_Radius}
\end{figure*}

We implemented a radius-dependent cut-off approach to resolve this problem, as illustrated in Fig.~\ref{fig:Prob_Radius}. The fitted King profile model, in this process, also plays a significant role. For each ring $i$, we determine a specific probability threshold $P_i$ such that the number of selected stars matches the number predicted by the King model ($N_{cl,i}$), as described in Equation \ref{eq:cluster_no}. Mathematically, the approach is
\begin{equation}
\label{eq:cluster_no}
   Np_{i}(P\geq P_i) \;\approx\; N_{cl,i}
\end{equation}
where $P_i$ is the probability in the $i$-th ring, giving the number of member stars as $Np_{i}$, which should match, as shown in Fig.~\ref{fig:Prob_Radius}, the number of stars from the King model, $N_{cl,i}$.
For example, if there are, as determined by the King fit, 100 member stars in shell $i$, the probability cut-off values $P_i$ will also result in the same number, 100 stars. The left panel of Fig. \ref{fig:Prob_Radius} illustrates the relationship between the probability threshold $P_i$ and the radial distance $r_i$. The indices of these member stars in ring $i$ in Python are provided as
\[ Index_i \;=\; (\; P >= P_i \; ) \]
where the probability values obtained from the pyUPMASK code, $P$, are denoted.

\textcolor{black}{The King density profile derived from the observed RDP serves as the primary structural benchmark for validating the membership separation method and constraining the total number of cluster members.} Conversely, an inaccurate membership method or an inappropriate probability cut-off may lead to significant overestimation or underestimation of the member count.

\begin{figure*} 
   \centering
   \includegraphics[width=0.65\linewidth]{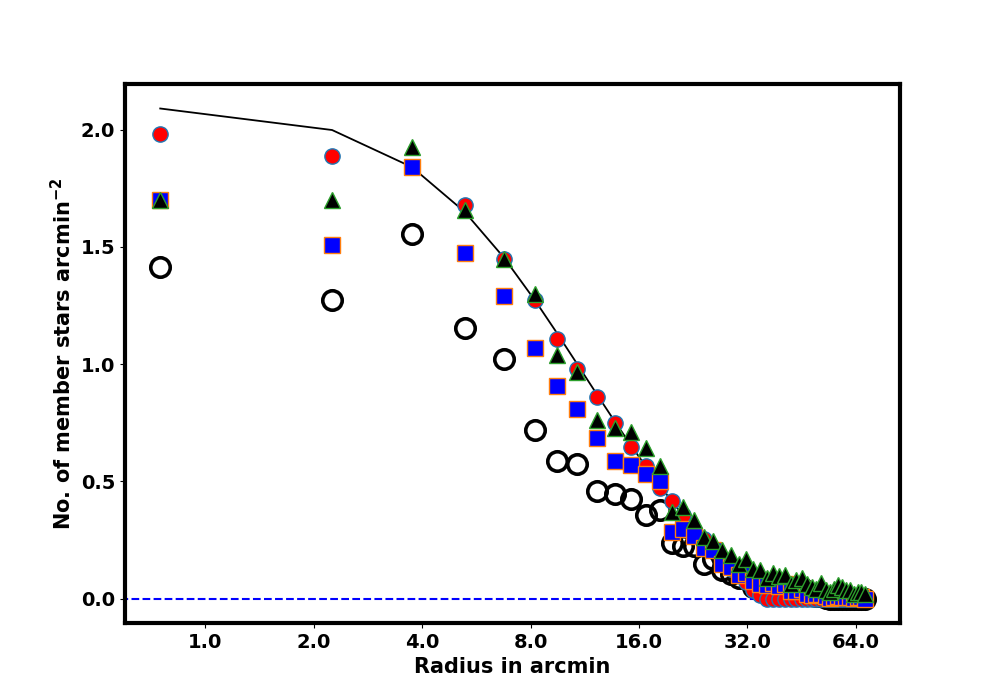}
   \caption{The density profile of member stars in the cluster is represented. The red dots are the  \textit{Gaia DR3} probable member stars density,  while the solid line illustrates the fitting of the King star density model for the cluster, derived from equation \ref{eq:king62}. The triangles are members from \cite{Groeningen2023} while the squares are members from \cite{hunt2024improving}. The open circles are members from \cite{Kos2024}. Neither of these groups follows the King model.}   
   \label{Fig:test_king}
   \end{figure*}  

\section{Photometric Properties of NGC 2168} \label{sec:phot}

In the context of OCs, to evaluate theoretical models of stellar evolution, color-magnitude diagrams (CMDs) use empirical isochrones \cite{Marigo2017, Spada2017}. CMDs serve as effective tools for estimating key cluster parameters such as distance, age, and metallicity. Additionally, valuable insights into the stellar masses within the cluster can be obtained by comparing observed CMDs with theoretical isochrones. The theoretical isochrones utilized in this research were obtained from CMD 3.9\footnote{http://stev.oapd.inaf.it/cgi-bin/cmd}, using PARSEC version 1.25S \citep{Bressan2012}.

\subsection{Extinction}
\label{sec:extin}
\textcolor{black}{A reliable interstellar extinction law is essential for deriving accurate photometric parameters. In this study, we adopted the procedure of \cite{Nasser2025a} and the method described by \cite{Wang2019} to compute extinction coefficients for the \textit{Gaia}, 2MASS, and optical bands using the general relation $A_{\lambda} = a A_V$.}

\textcolor{black}{TFor the \textit{Gaia} bands, we adopted $A_G/A_V = 0.789$, $A_{BP}/A_V = 1.002$, and $A_{RP}/A_V = 0.589$. For the near-infrared 2MASS bands, the corresponding values are $A_J/A_V = 0.243$ and $A_{K_s}/A_V = 0.078$. In the optical regime, we followed the extinction law of \cite{Cardelli1989} and \cite{Donnell1994}, assuming $R_V = 3.1$.}

\begin{figure}[htbp]
    \centering
    \begin{minipage}[t]{0.4\textwidth}
        \centering 
        \includegraphics[width=\linewidth]{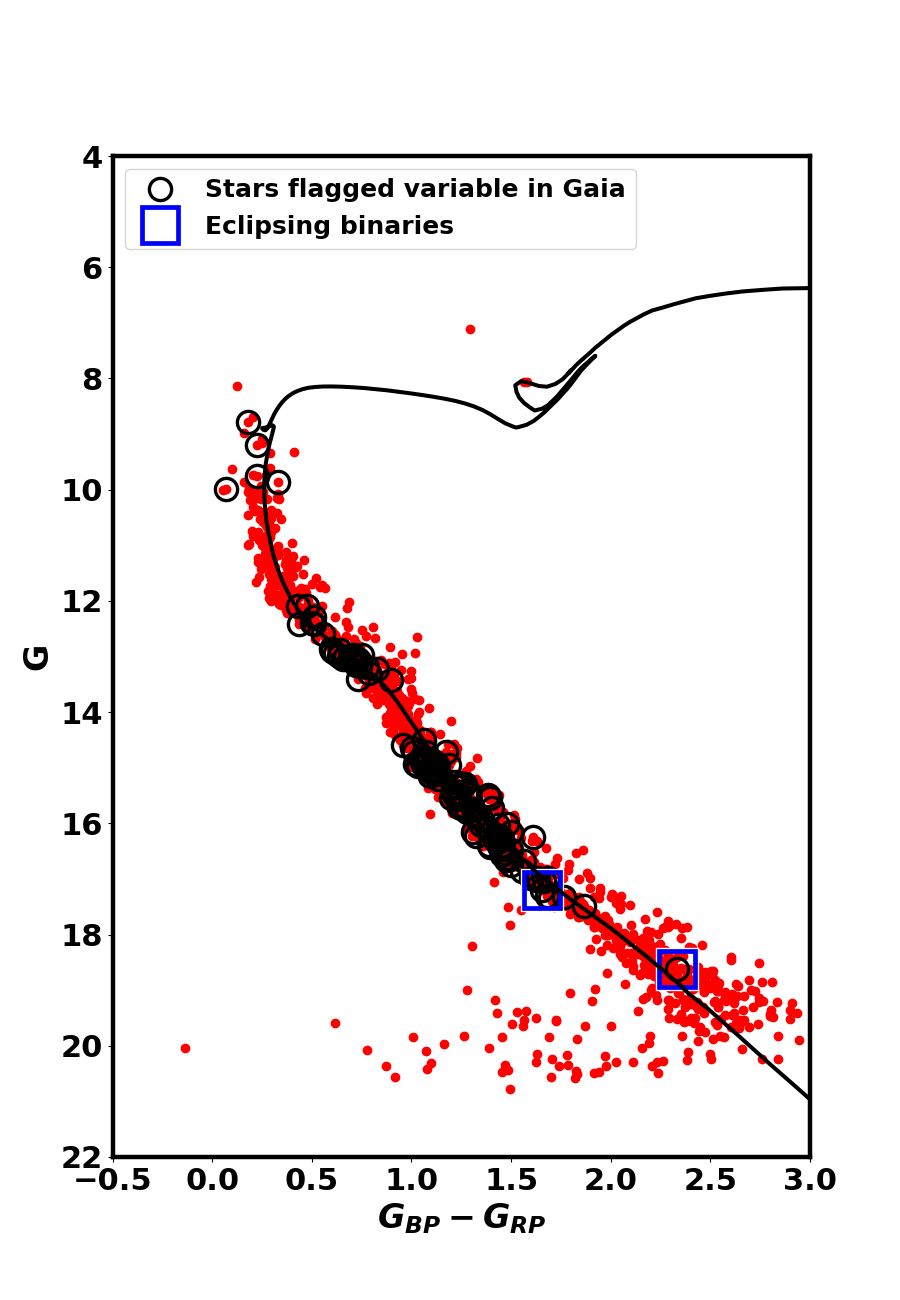}
        \caption{Color--magnitude diagram (CMD) of NGC~2168. The best-fitting isochrone corresponds to an age of 190~Myr, a metallicity of $Z = 0.0133$, and [M/H] $= -0.0479$~dex. Open circles denote cluster members classified as variable stars in the \textit{Gaia} DR3 archive, while square symbols indicate eclipsing binary systems identified by \citet{GaiaCollaboration2023} (VizieR catalogue I/358).}
        \label{fig:gaia_cmd}
    \end{minipage}
    \hfill 

    \begin{minipage}[t]{0.4\textwidth}
        \centering
        \includegraphics[width=\linewidth]{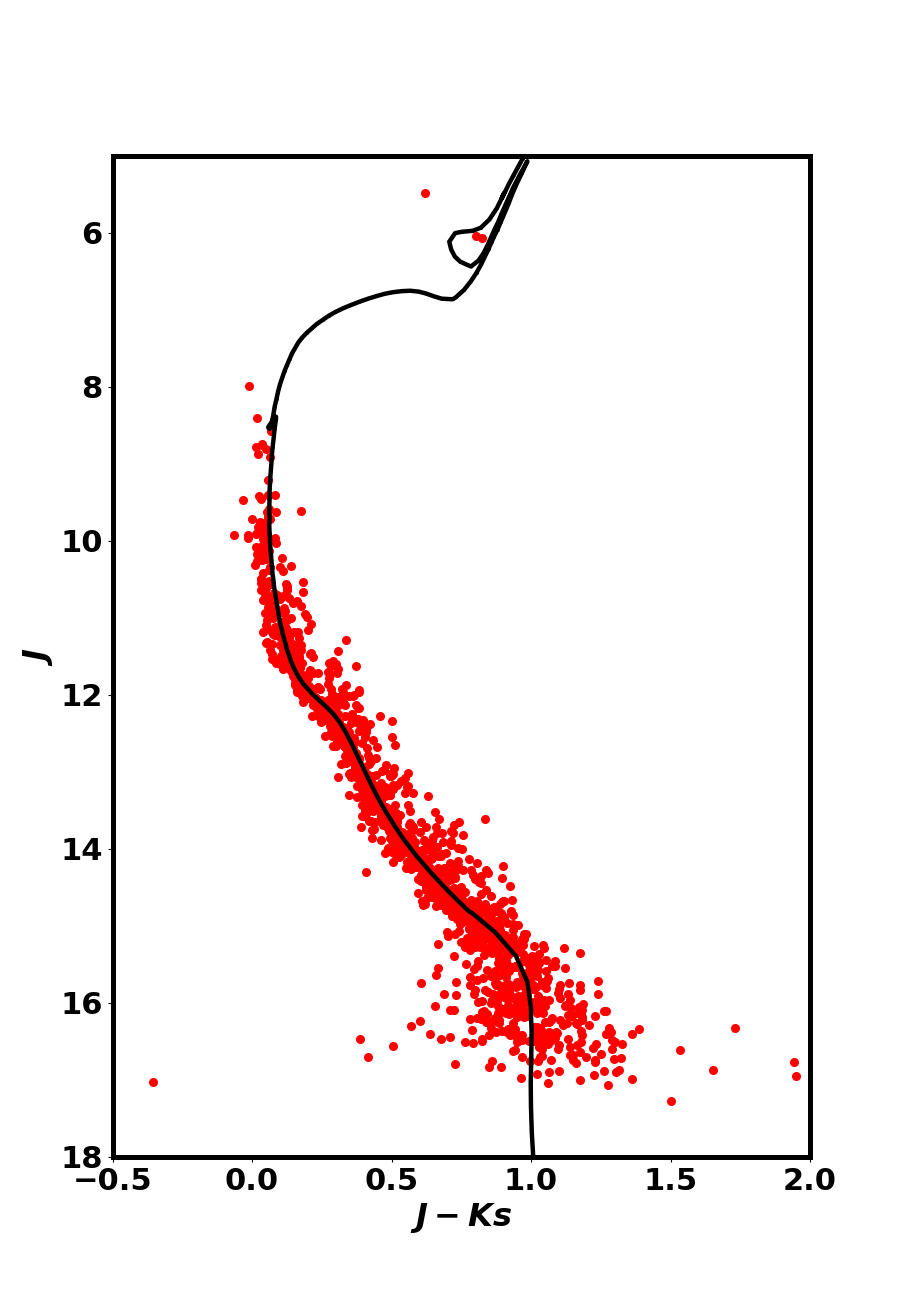} 
        \caption{Near-infrared color--magnitude diagram constructed from 2MASS photometry. The same isochrone adopted for the \textit{Gaia} CMD is overplotted. The derived intrinsic distance modulus and color excess are $(m-M)_0 = 9.6 \pm 0.23$~mag and $E(J-K_s) = 0.15 \pm 0.02$~mag, respectively.}
        \label{fig:cmd_2mass}
    \end{minipage}
\end{figure}

\textcolor{black}{TUsing these coefficients, the extinction–color excess relations were derived. In particular,
\begin{align}
\begin{split}
A_G &= 1.88 \, E(G_{BP} - G_{RP}), \\
A_V &= 3.1 \, E(B-V),
\end{split}
\end{align}
and the color excess ratios were expressed as
\begin{align} \label{eq:Gaia_BV_ratio}
E(G_{BP} - G_{RP}) = 1.29 \, E(B-V).
\end{align}}
\textcolor{black}{Through isochrone fitting, we derived the color excess and the corresponding extinction. The intrinsic distance modulus was obtained by correcting the observed modulus for extinction, i.e. $(m-M)_0 = (m-M)_{\mathrm{obs}} - A_{\lambda}$, where $A_{\lambda}$ denotes the extinction in the relevant passband. The cluster distance was then calculated using the standard distance–modulus relation. Using the \textit{Gaia} DR3 photometry, the CMD of NGC 2168 is presented in Fig.~\ref{fig:gaia_cmd}, together with the adopted theoretical isochrones from \cite{Marigo2017}.}

\textcolor{black}{The intrinsic distance modulus and the color excess are found to be $9.65 \pm 0.12$ mag and $0.40 \pm 0.04$ mag, respectively. These values correspond to an isochrone-based distance of $850 \pm 62$ pc. The mean cluster parallax of $1.153 \pm 0.052$ mas yields a geometric distance of $867 \pm 39$ pc, which is fully consistent with the isochrone determination within the uncertainties. This agreement supports the reliability of the adopted photometric solution. For completeness, we note that the Bayesian distances reported by \cite{Bailer-Jones2021} are also compatible; however, their exponentially decreasing space-density prior is optimized for field stars and may not strictly represent the spatial distribution of cluster members.}

\begin{figure*}[ht]
\centering
\includegraphics[width=.65\linewidth]{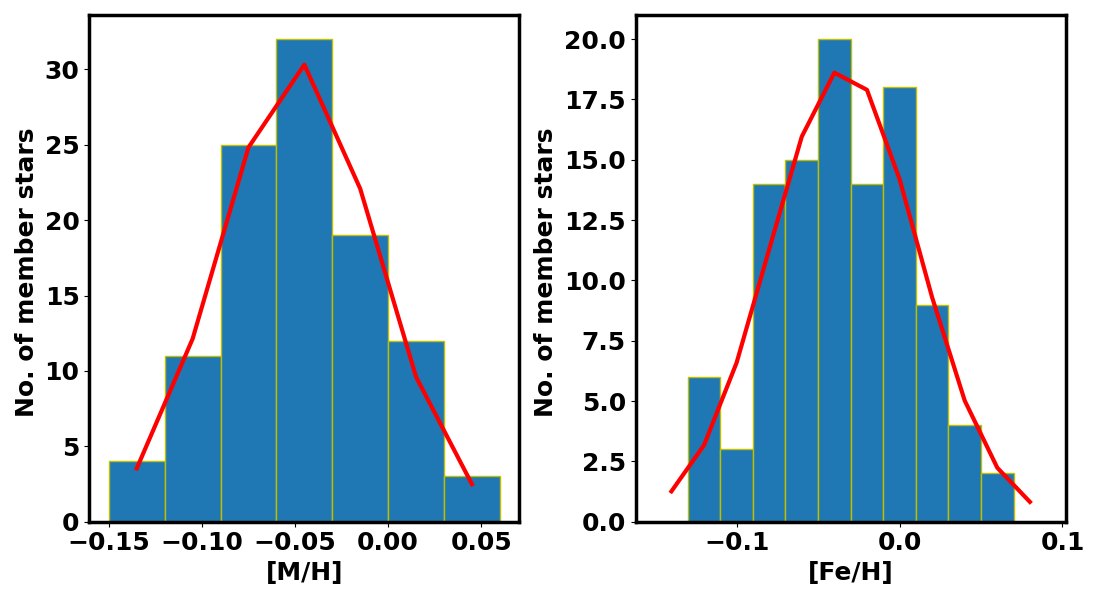}
\caption{Left panel: histogram of [M/H] values for cluster members cross-matched with the catalogue of \citet{Abdurro2022} (VizieR catalogue III/286), together with the best-fitting Gaussian distribution, yielding a mean metallicity of [M/H] $= -0.048$~dex. Right panel: histogram of [Fe/H] values for the same sample, with a Gaussian fit indicating a mean value of [Fe/H] $= -0.033$~dex. For comparison, the metallicity adopted for the photometric isochrone fitting is [M/H] $= -0.0479$~dex.}
\label{fig:metal}
\end{figure*}

The fitted isochrone indicates a cluster age of 190$\pm$ 12 Myr, with a metallicity of $Z$ = 0.0133 ([M/H]= -0.0479 dex) Moreover, we derived the [Fe/H] value using the relation provided by \citet{Bovy2015}\footnote{https://github.com/jobovy/isodist/blob/main/isodist/Isochrone.py}
\begin{equation}
z_x = 10^{ [Fe/H] \;\;+\;\; \log{\left( \dfrac{z_\odot}{ 0.752 -  2.78*z_\odot } \right)} }
\end{equation}
and 
\begin{equation}
    Z = \dfrac{(\; 0.7515*z_x \;)}{( \; 2.78*zx+1.0 \;)}
\end{equation}
Assuming a solar metallicity of $z_\odot = 0.0152$, we derive a metallicity of [Fe/H] = $-0.039$~dex, in good agreement with our previously obtained result. This derivation follows the same algorithm as used in previous studies \citep[e.g.,][]{Tasdemir23,Cinar2024}, ensuring consistency with established methods. In addition, we cross-matched our cluster members with the spectroscopic catalogue of \citet{Abdurro2022}. This comparison yielded 257 common member stars. From a Gaussian fit to the metallicity distributions, we obtain mean values of [M/H] = $-0.048$ and [Fe/H] = $-0.033$ (see Fig.~\ref{fig:metal}). These spectroscopic estimates are fully consistent with our photometric metallicity within the quoted uncertainties. In contrast, our results differ from those reported by \citet{Reddy2019}, who derived their mean metallicity from only two giant stars, resulting in a less statistically representative value.

We further cross-matched our member sample with the 2MASS catalogue to construct an independent near-infrared color--magnitude diagram (CMD), shown in Fig.~\ref{fig:cmd_2mass}. From this CMD, we derive an intrinsic distance modulus of $(m-M)_0 = 9.6 \pm 0.23$~mag and a color excess of $E(J-K_s) = 0.15 \pm 0.02$~mag. Both values are in excellent agreement with those inferred from \textit{Gaia} photometry. We also find that the ratio $E(G_{\rm BP}-G_{\rm RP})/E(J-K_s)=2.66$, and adopting $E(G_{\rm BP}-G_{\rm RP})=0.40$ together with the conversion $E(G_{\rm BP}-G_{\rm RP}) \simeq 1.29\,E(B-V)$, we obtain $E(B-V)\approx 0.31$~mag, consistent with the relation given in Equation~\ref{eq:Gaia_BV_ratio} and the literature compilation in Table~\ref{tab:literature}

\begin{figure}[ht]
\centering
\includegraphics[width=\linewidth]{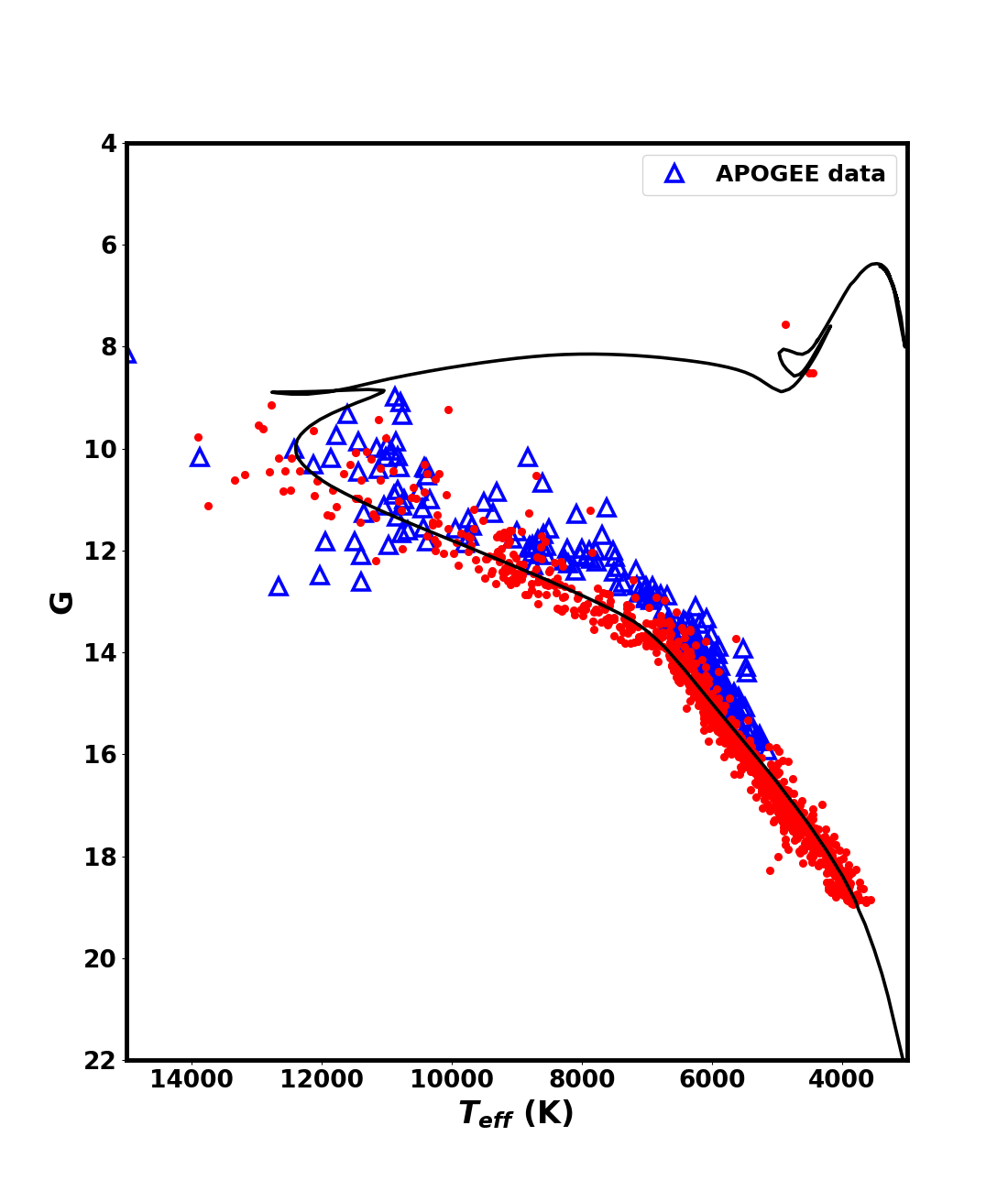}
\caption{$G$ magnitude versus effective temperature ($T_{\rm eff}$) for cluster members. 
Effective temperatures are adopted from \citet{Anders2022}. Triangle symbols indicate stars cross-matched with the catalogue of \citet{Abdurro2022}, which show good agreement with the main data set. The Abdurro et al.\ (2022) parameters were obtained from the VizieR catalogue (III/286).}
\label{fig:G_temp}
\end{figure}

\begin{figure}[ht]
\centering
\includegraphics[width=.95\linewidth]{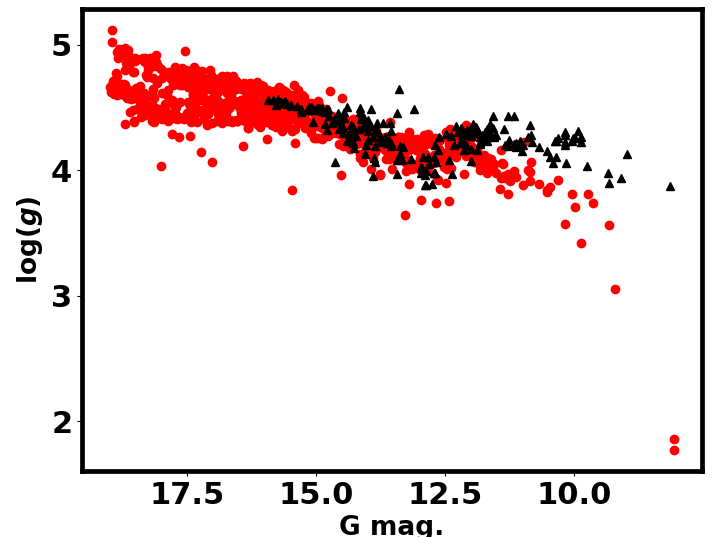}
\caption{$G$ magnitude versus the logarithm of the surface gravity ($\log g$) from \textit{Gaia} DR3. Two cluster members exhibit low surface gravity values, confirming their classification as giant stars. Triangle symbols indicate stars cross-matched with the catalogue of \citet{Abdurro2022}, which are in good agreement with the \textit{Gaia} DR3 measurements.} 
\label{fig:surf_gravity}
\end{figure}

Effective temperature is a key physical parameter in stellar population analyses, particularly when reliable measurements are available. We therefore adopted effective temperatures from the catalogue of \citet{Anders2022} and examined their distribution as a function of $G$ magnitude (Fig.~\ref{fig:G_temp}). The overplotted solid line represents the same isochrone used in the CMD analysis. This comparison provides an important consistency check for the isochrone fitting and aids in the identification and classification of evolved stellar populations. In particular, three giant stars are clearly identified in both diagrams, in agreement with their low surface gravity values shown in Fig.~\ref{fig:surf_gravity}. Furthermore, we identified 125 member stars marked as variable in the Gaia archive, as  shown in Fig. \ref{fig:gaia_cmd}.  Among these, two are categorized as eclipsing binaries \citet{GaiaCollaboration2023}, as denoted by the square symbols in the same figure.  The details concerning these stars are not within the scope of this paper.

%

%
\section{Cluster Mass, Mass Function and Mass Populations}\label{sec:mass}
The determination of individual stellar masses is a prerequisite for analyzing the cluster's dynamical state and mass segregation. While the absolute magnitude ($M_{\rm G}$) and dereddened intrinsic color index ($(G_{\rm BP}-G_{\rm RP})_0$) derived from the fitting of an isochrone provide the basis for this calculation, translating these parameters into stellar mass requires a robust methodology. Simple polynomial relations often fail to capture the curvature of evolutionary tracks, particularly around the Main Sequence Turn-Off (MSTO) and the sub-giant branch, where the mass-luminosity relation becomes non-linear.

To address this, we implemented a two-dimensional interpolation scheme using the \textit{SmoothBivariateSpline} routine from the \textsc{SciPy} library \cite{SciPy2020}. This method treats stellar mass ($M_{\star}$) as a continuous function of both magnitude and color, i.e., $M_{\star} = f(M_{\rm G}$, $(G_{\rm BP}-G_{\rm RP})_0)$, ensuring precise mapping across the entire Main Sequence.

\begin{figure}[ht]
\centering
\small
\includegraphics[width=8.2cm, angle=0]{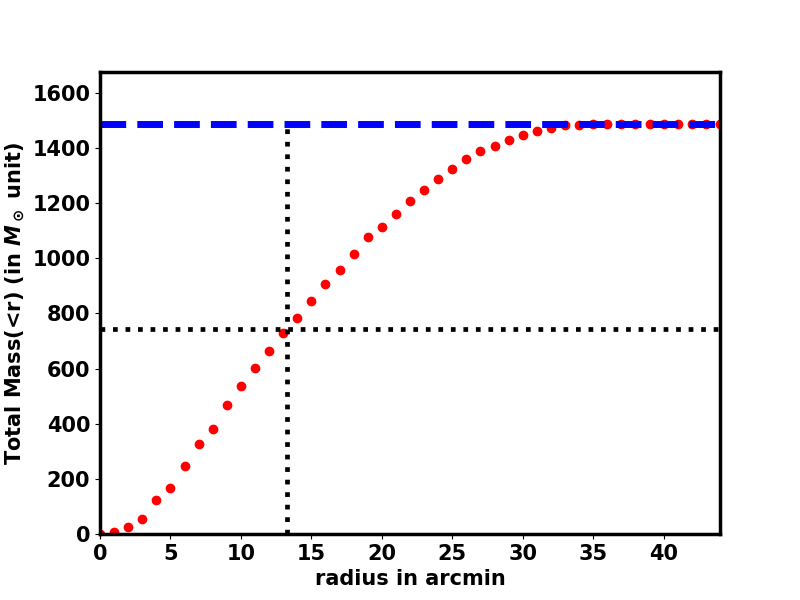}
\caption{The mass profile $M(<r)$ of NGC 2168. The horizontal blue dashed line indicates the total mass, while the yellow dashed line represents the half-mass radius, $R_h$.} 
\label{fig:mass-profile}
\end{figure}

By integrating the masses of all high-probability members, we derived a total cluster mass of $\mathcal{M}_{\rm cl} = 1485\pm$59$  M_{\odot}$. Furthermore, we constructed the cumulative radial mass distribution (Fig.~\ref{fig:mass-profile}) to determine the half-mass radius, $R_h = 13.26$ arcmin, corresponding to 3.24 pc. This structural parameter, defining the radius containing 50\% of the total mass, is a critical input for calculating the dynamical relaxation time, $T_R$:
\begin{equation}
T_R = \dfrac{8.9 \times 10^5 \sqrt{N} \times R_{h}^{1.5}}{\sqrt{\bar{m}} \times \log(0.4N)} \label{eq:relax_time}
\end{equation}

\noindent where $N$ represents the number of member stars included in the mass summation and $\bar{m}$ ($1.22\;M_{\odot}$) is the mean stellar mass. We calculated the relaxation time ($T_R$) for NGC 2168 as 65.4 $\pm$12.5 Myr, which is considerably shorter than the age of the cluster (190 Myr), indicating that the system is dynamically relaxed.

%
%
\subsection{The Traditional Salpeter Mass Function}
\begin{figure}
\centering
\includegraphics[width=8.cm, angle=0]{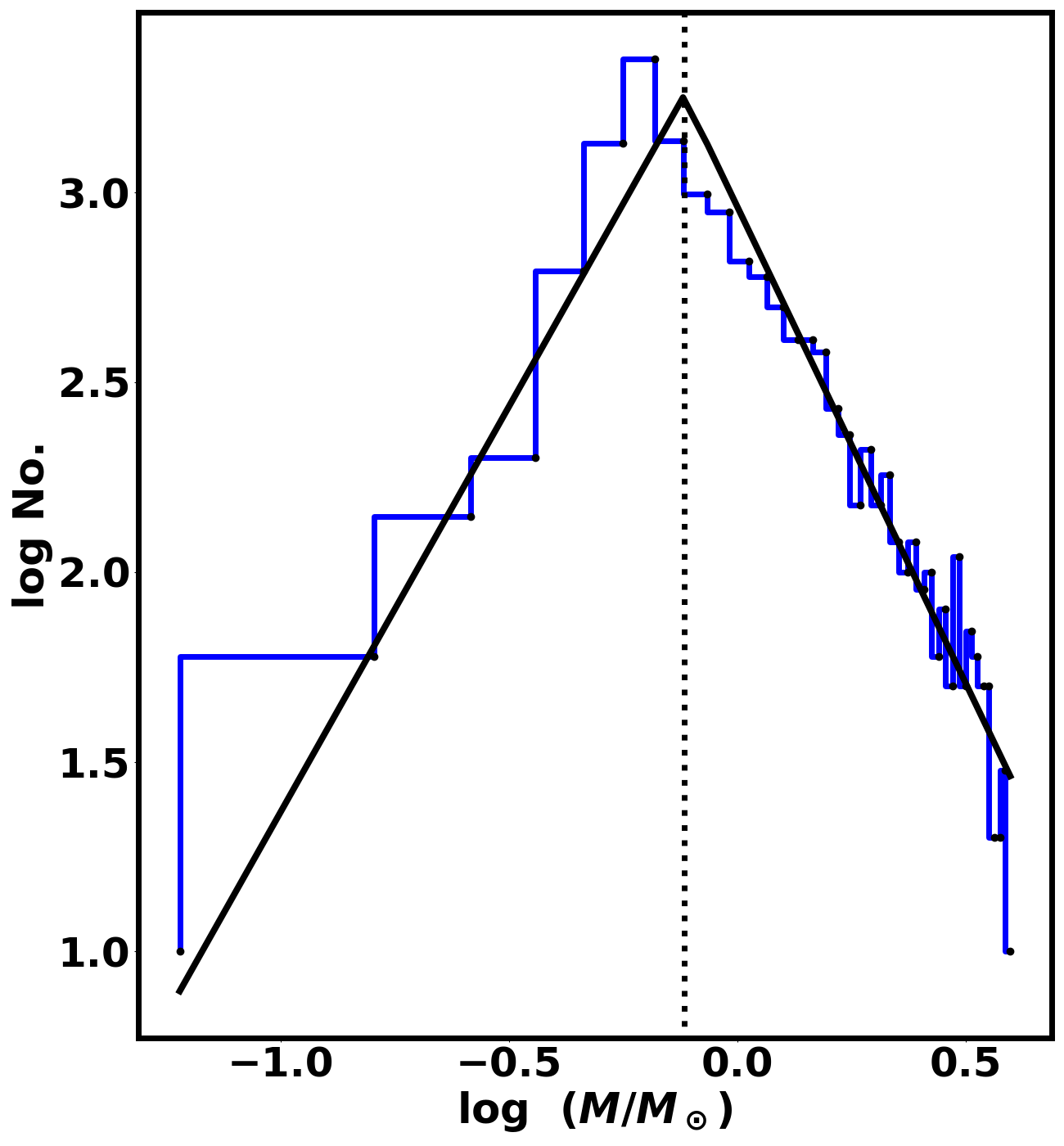}
\caption{Salpeter mass function \citep{Salpeter1955}.}
\label{fig:MF2}
\end{figure}

In our earlier publications, \citet{Nasser2025a} and \citet{Nasser2025b}, we represent the mass function (MF) using a two-segment (broken) power-law form as demonstrated by the work of \cite{Almeida2023}. This approach stands in contrast to the single power law equation that was originally proposed by \cite{Salpeter1955}. The broken power-law parametrization allows the low- and high-mass regimes to be described independently via slopes $-\alpha_1$ and $-\alpha_2$ separated by a critical mass $M_{cr}$. It can be represented as follows
\begin{equation} 
f(M) = \frac{dN}{dm}  = 
\Bigg \{
\begin{array}{lcc} 
      K_1 \;  m^{-\alpha_1}  &, & \text{if } m \le M_{cr}\\
      K_2 \; m^{-\alpha_2}  &, & \text{if } m >  M_{cr}
\end{array}
\label{eq:Salpeter_mf}    
\end{equation}
under the condition that the function $f(M)$ is continuous:
\begin{equation*}
K_1 \; M_{cr}^{-\alpha_1} = K_2\; M_{cr}^{-\alpha_2}
\end{equation*}
where {dN}/{dm} indicates the number of stars within the mass range $m$ to $m + dm$, see Fig. \ref{fig:MF2}. The parameters $\alpha_1$ and $\alpha_2$ represent the low and high mass slopes of the mass function, respectively, while $M_{cr}$ marks the critical mass where the slope changes between the two regimes. In many open clusters, the high mass slope $\alpha_2$ is near to \citet{Salpeter1955} value of 2.35.  The fitting of the primary population is performed using the curve\_fit function of the Scipy python package with $\alpha_1$, $\alpha_2$, $M_{cr}$, $K_1$, and $K_2$ treated as free parameters, constrained by the condition $ f(M^{-}_{cr}) = f(M^{+}_{cr})$, within the mass range of 0.05 to 4.0 $M_{\odot}$, see Table \ref{tab:mass_func}. The high mass slope $\alpha_2$ is found to be 2.37, which corresponds with the results of \citet{Salpeter1955} value (2.35), see all resuts in Table \ref{tab:mass_func}.\\

\begin{table}[t]
\footnotesize
\centering
\setlength{\tabcolsep}{3pt} 
\renewcommand{\arraystretch}{1.0}
\caption{Parameters of the two--power-law fit to the cluster mass function, see equation~\ref{eq:Salpeter_mf}.}
\label{tab:mass_func}
\begin{tabular}{ccccc}
\hline\hline
$K_1$ & $K_2$ & $M_{\rm cr}$ ($M_\odot$) & $\alpha_1$ & $\alpha_2$ \\
\hline
$3.40 \pm 0.21$ & $2.93 \pm 0.08$ & $0.73 \pm 0.13$ & $-2.20 \pm 0.13$ & $2.37 \pm 3.15$ \\
\hline
\end{tabular}
\end{table}

%
%
\subsection{Gaussian Representation of the Present-Day Mass Distribution}

\begin{figure}
\centering
\includegraphics[width=8.cm] {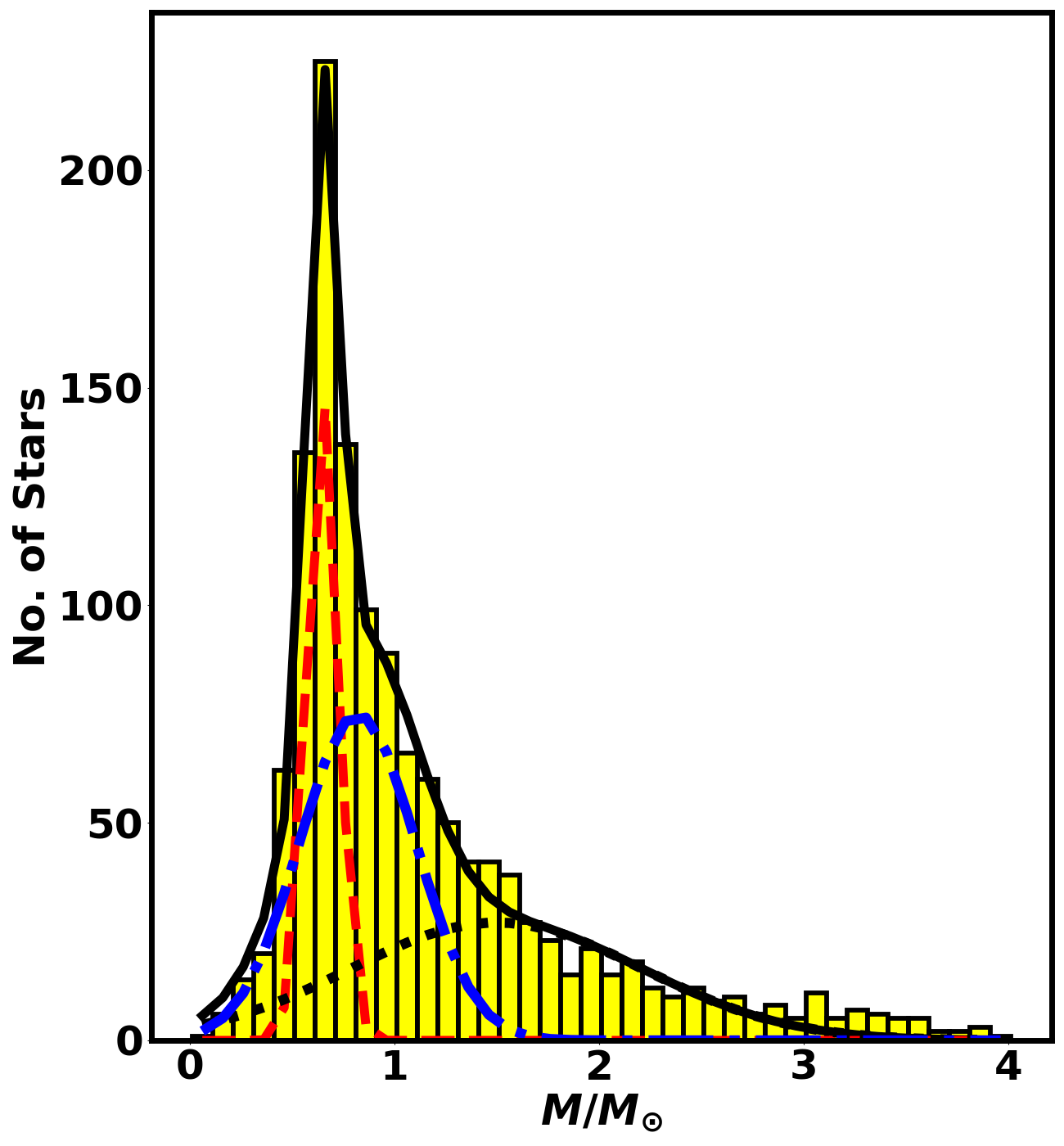}
\caption{The mass distribution for NGC 2168 is represented by a fitting of three Gaussian functions.}
\label{fig:MF1}
\end{figure}

\textcolor{black}{In addition to the traditional power-law parametrization, we explored an alternative functional description to examine whether the observed present-day mass distribution exhibits multimodal structure. Following our earlier works \citep{Nasser2025a,Nasser2025b}, we fitted the mass histogram using a sum of Gaussian components,
\begin{equation}\label{eq:gass_mf}
\frac{dN(m)}{dm} = \sum_i N_{0,i} \exp\left(-\frac{(m-\mu_{m,i})^2}{2\sigma_{m,i}^2}\right).
\end{equation}
The observed distribution is statistically reproduced by a three-component Gaussian mixture (Fig.~\ref{fig:MF1}; parameters listed in Table~\ref{tab:mass_dist}). We emphasize that this representation is purely phenomenological and is not intended as a physical alternative to the classical (broken) power-law initial mass function. Multimodality may arise from observational incompleteness, unresolved binaries, or residual contamination. Therefore, the Gaussian components are interpreted only as a descriptive tool for the present-day mass distribution.}

\begin{table}[t]
\footnotesize
\centering
\caption{Parameters of the three--Gaussian fit to the cluster mass function, see equation \ref{eq:gass_mf}.}
\label{tab:mass_dist}
\begin{tabular}{cccc}
\hline\hline
Component & $N_0$ (stars) & $\mu_m$ ($M_\odot$) & $\sigma_m$ ($M_\odot$) \\
\hline
1 & $147.4 \pm 13.7$ & $0.65 \pm 0.05$ & $0.077 \pm 0.003$ \\
2 & $75.0 \pm 8.3$   & $0.82 \pm 0.07$ & $0.28 \pm 0.02$ \\
3 & $27.1 \pm 4.6$   & $1.50 \pm 0.04$ & $0.56 \pm 0.08$ \\
\hline
\label{ltab:gass_MF}
\end{tabular}
\end{table}

\subsection{Stellar Populations}
Although the high-mass slope $\alpha_2$ derived in Equation \ref{eq:Salpeter_mf} is consistent with the canonical Salpeter value, relying solely on a power-law approximation provides an incomplete picture of the cluster's complex dynamical state.
Stellar populations provide key constraints on star-cluster formation and subsequent evolution. In this study, however, we use the term \emph{mass populations} exclusively to denote distinct components in the fitted stellar mass distribution (e.g., the individual components of the multi-Gaussian model). This terminology should not be confused with \emph{multiple stellar populations} in the classical sense, which refers to stars with different ages and/or chemical abundance patterns.

\textcolor{black}{In our previous analysis of the old, massive open cluster NGC~6791 \citep{Nasser2025d}, the present-day mass distribution was statistically reproduced by two Gaussian components. In that context, the term multiple mass components was introduced to describe the fitted structure of the mass distribution, without implying distinct stellar generations in age or chemical composition.}

\textcolor{black}{Within the same phenomenological framework, the mass distribution of NGC~2168 is statistically characterized by three Gaussian components. In the following section, we discuss these components cautiously in relation to possible dynamical effects and early cluster evolution, emphasizing that they represent a descriptive fit to the present-day mass distribution rather than a physically distinct set of stellar populations.}

\subsection{The Initial Dynamics}

The Gaussian representation of the present-day mass distribution, and the resulting identification of multiple mass populations, can provide complementary insight into the early dynamical state of a cluster and its subsequent evolution within the Galactic environment.

The analysis reveals that the mass distribution of NGC 2168 is best described by three distinct Gaussian components. This multimodal structure suggests a complex formation history or dynamical evolution, potentially hinting that the cluster may have formed from the merger of distinct stellar clumps or experienced significant dynamical sub-structuring. While a merger scenario is a compelling explanation for these distinct populations, further chemical tagging is required to fully confirm this hypothesis.

In comparison, the NGC 6791 cluster is a single cluster with two mass populations. This suggests that they were initially two clusters that were brought together by the gravitational potential and are located far from any disruptive forces.

As previously mentioned, the Gaussian mass function or mass populations serves as a crucial tool for investigating the initial conditions of any binary or any system of clusters, or even a single cluster, and is considered to be more realistic and accurate than the Salpeter function. The primordial binary star cluster requires a reassessment through the application of a Gaussian function to ascertain if they were previously a singular system that was torn apart by gravitational tides, or if they were binary from their inception. This approach enables us to investigate not only the dynamics of these clusters but also the dynamics of the Galaxy.
%
%
%
%
\section{Cluster Kinematics and Dynamics}
OCs are outstanding markers for tracing the evolution of the Galactic disc. The release of \textit{Gaia DR3} allows for the investigation of their  Dynamics and Kinematics with an unprecedented level of precision and accuracy.
  The center of the cluster is located at 92.26$\pm$0.1 and 24.30$\pm$0.1, which corresponds to the Galactic coordinates l = 186.64$\pm$0.1 and b=2.21$\pm$0.1.
\subsection{Astrometric Parameters and Kinematic Structure}
\label{sec:distance}

\begin{figure*}
\centering
\includegraphics[width=0.99\linewidth]{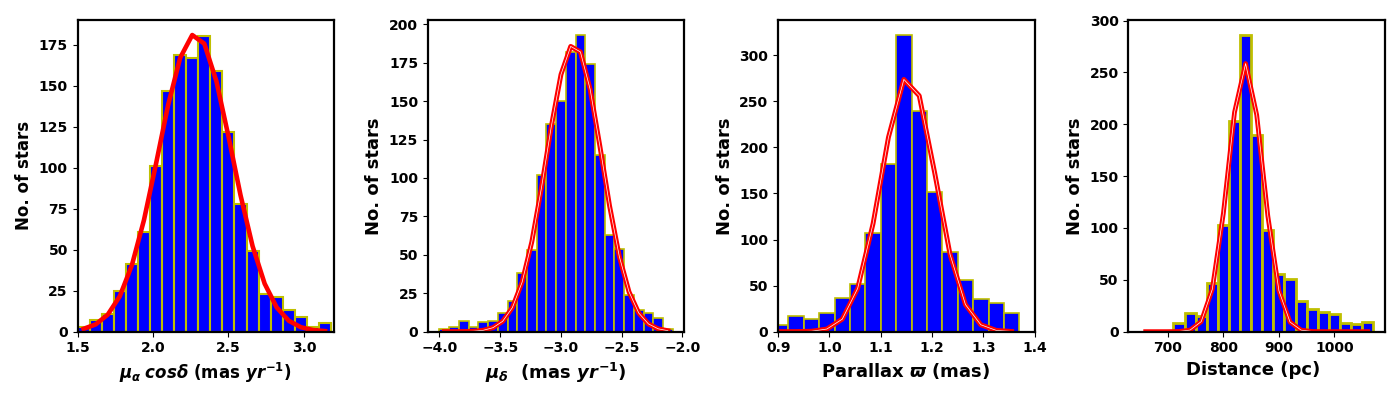}
\caption{The members' proper motions, parallaxes, and distances histograms. The solid red lines are Gaussian fits.}
\label{Fig:param_fit}
\end{figure*}

The distributions of the proper-motion components and trigonometric parallaxes for the selected cluster members are well described by Gaussian profiles, as shown in Fig.~\ref{Fig:param_fit}. The Gaussian fits yield mean proper motions of $\mu_{\alpha}\cos\delta = 2.28$ mas yr$^{-1}$ and $\mu_{\delta} = -2.89$ mas yr$^{-1}$, \textcolor{black}{with dispersions of 0.24 and 0.23 mas yr$^{-1}$, respectively. Given the number of members ($N = 1397$), the uncertainties of the mean values are significantly smaller, amounting to $\sim$0.006 mas yr$^{-1}$ in each component.}

To improve the accuracy of the trigonometric parallax measurements, we apply the \textit{Gaia} DR3 parallax zero-point correction following the prescription of \citet{Lindegren2021}, implemented via the \texttt{gaiadr3\_zeropoint} Python module. A Gaussian fit to the corrected parallax distribution yields a mean parallax of $\varpi = 1.154 \pm 0.052$~mas (Fig.~\ref{Fig:param_fit}). All derived astrometric parameters are summarized in Table~\ref{tab:center}.

Although trigonometric parallaxes provide the most direct observational constraint on stellar distances, they cannot be trivially inverted to obtain reliable distances, owing to the nonlinear transformation and the impact of measurement uncertainties, particularly for distant stars. Small parallax uncertainties can result in large uncertainties in the inferred distances, and negative parallax values prevent direct inversion. For these reasons, distance estimates based on explicit probabilistic approaches are preferred.

We therefore adopt the distance estimates provided by \citet{Bailer-Jones2021}, who derived probabilistic distances for 1.47 billion stars in the \textit{Gaia} EDR3 catalogue. For our cluster members, the resulting distance distribution is again well represented by a Gaussian profile. The mean cluster distance is found to be $840 \pm 54$~pc (Fig.~\ref{Fig:param_fit}), in good agreement with the photometric distance derived independently within the quoted uncertainties.

\begin{figure}[htbp]
    \centering
    \begin{minipage}[t]{0.32\textwidth} 
        \centering
        \includegraphics[width=\linewidth]{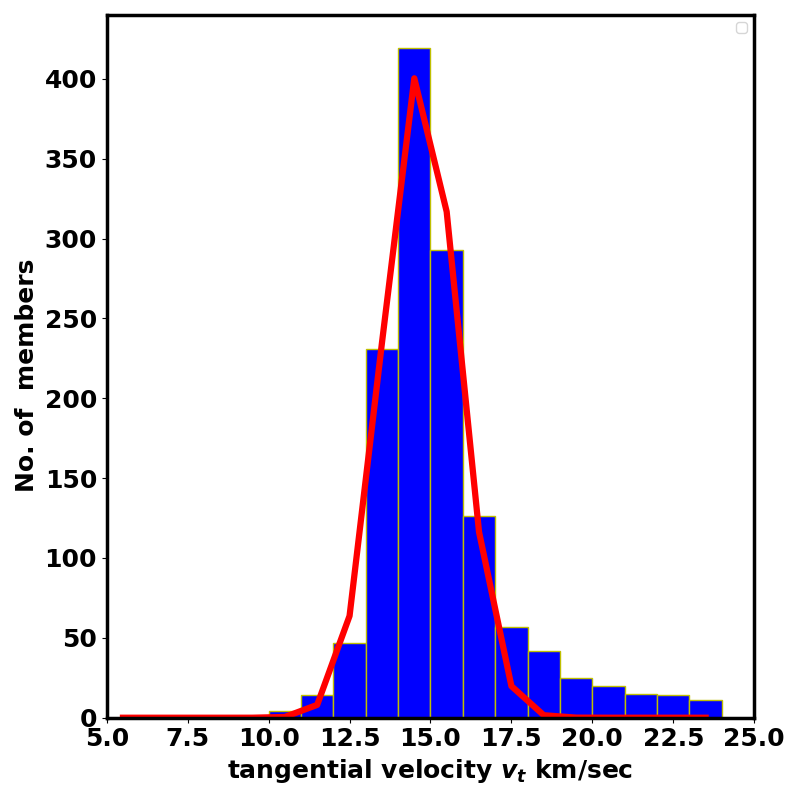} 
        \caption{The tangential velocities are represented in a histogram. The red solid line indicates the Gaussian fit, which has a mean value of 14.7 $km\;s^{-1}$.}
        \label{fig:vt}
    \end{minipage}
    \hfill 
    \begin{minipage}[t]{0.32\textwidth} 
        \centering
        \includegraphics[width=\linewidth]{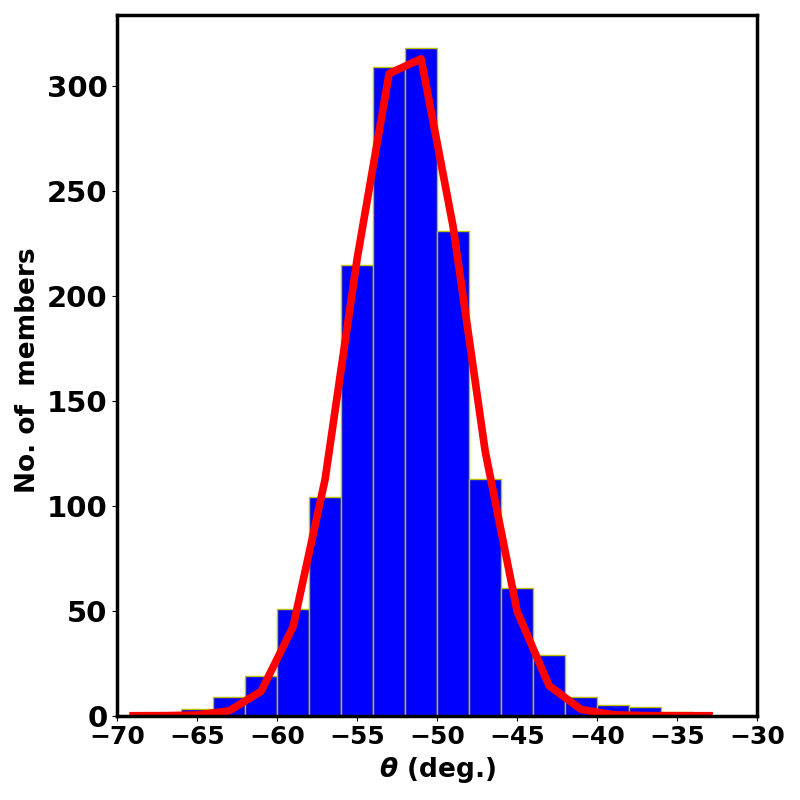} 
        \caption{The histogram illustrating the distribution of theta members is presented. The solid red line represents the Gaussian fit with a mean of -51.85$^{\circ}$.}
        \label{fig:theta}
    \end{minipage}
    \begin{minipage}[t]{0.32\textwidth} 
        \centering
        \includegraphics[width=\linewidth]{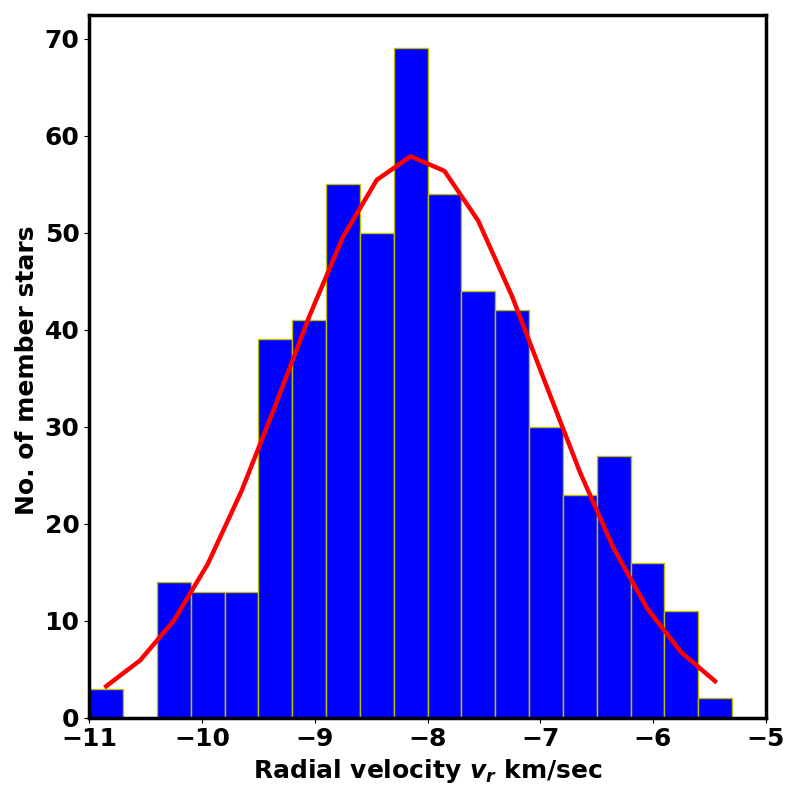} 
        \caption{The histogram of member radial velocities is derived from the data presented in \cite{Leiner2015}. The red solid line represents the Gaussian fit with a mean of -8.16 $km\;s^{-1}$}
        \label{fig:rv}
    \end{minipage}
\end{figure}

Stars in an cluster are expected to share similar space motions. The tangential velocity of each star can be computed from its total proper motion,
$\mu = \sqrt{(\mu_{\alpha}\cos\delta)^2 + \mu_{\delta}^2}$,
and distance $d$. The tangential velocity in km~s$^{-1}$ is given by
\begin{equation}
v_t = 4.74 \, \mu \, d ,
\end{equation}
where $\mu$ is expressed in arcsec~yr$^{-1}$ and $d$ in parsecs. The numerical factor 4.74 arises from the conversion between angular motion and linear velocity. The resulting tangential velocity distribution for NGC~2168 is shown in Fig.~\ref{fig:vt} and exhibits a nearly Gaussian shape, with a mean value of $14.7 \pm 0.8$~km~s$^{-1}$.

Because cluster members move through space along nearly parallel trajectories, their proper motion vectors appear to converge toward a common point on the sky, known as the convergent point. Consequently, the tangential velocity alone does not fully characterize the kinematic structure of the cluster. An additional diagnostic is the angle $\theta$, which describes the direction of motion in the $(\mu_{\alpha}\cos\delta,\,\mu_{\delta})$ plane and is defined as
\begin{equation}
\theta = \tan^{-1}\left(\frac{\mu_{\delta}}{\mu_{\alpha}\cos\delta}\right).
\end{equation}
The distribution of $\theta$ for the cluster members is presented in Fig.~\ref{fig:theta}, yielding a mean value of $-51.9 \pm 2.4^{\circ}$. The observed dispersion in $\theta$ may reflect the dynamical state of the cluster, including its age and degree of gravitational binding, and also provides an independent consistency check on the adopted membership selection.

Accurate estimates of the space motion and orbital parameters of NGC 2168 rely critically on a robust determination of its mean radial velocity. To achieve this, we did not limit our analysis to the radial velocity measurements provided by the \textit{Gaia} DR3 catalogue \citep{GaiaCollaboration2023}. Instead, we complemented these data with additional radial velocity measurements compiled by \citet{2022A&A...659A..95T}, which incorporate large spectroscopic surveys such as \textit{APOGEE} and \textit{LAMOST}.

Radial velocity measurements were obtained by cross-matching our list of NGC 2168 members with the combined spectroscopic catalogues. After applying basic quality checks to ensure reliable measurements, a total of 368 member stars were retained for the final analysis. This sample consists of 7 stars from \textit{APOGEE}, 80 stars from \textit{LAMOST}, and 281 stars from the \textit{Gaia} DR3 catalogue, providing a statistically robust basis for the determination of the cluster radial velocity.

\begin{table*}[ht]
\caption{Representative sample of radial velocity measurements for NGC~2168.
The full table containing all 368 stars is available in the electronic version of the journal.}
\label{tab:rv_sample}
\centering
\footnotesize
\begin{tabular}{l l c c c c l}
\hline\hline
Order & $Gaia$ DR3 ID & $\alpha$ (deg) & $\delta$ (deg) & $V_{\rm R}$ (km s$^{-1}$) & $\sigma_{V_{\rm R}}$ (km s$^{-1}$) & Survey \\
\hline
1 & 3426290609695576064 & 92.4477 & 24.3857 & -5.16 & 1.35 & \textit{APOGEE}\\ 
2 & 3426270406169408512 & 92.3157 & 24.3788 & -8.67 & 0.60 & \textit{APOGEE}\\ 
3 & 3426307583406596608 & 92.3315 & 24.5285 & -7.37 & 0.47 & \textit{APOGEE}\\ 
4 & 3426290751431315456 & 92.4029 & 24.3722 & -7.36 & 4.03 & \textit{APOGEE}\\ 
5 & 3426265222145752064 & 92.3629 & 24.2292 & -15.74 & 0.84 & \textit{APOGEE}\\ 
6 & 3426311569136380800 & 92.2136 & 24.5920 & -7.65 & 0.70 & \textit{APOGEE}\\ 
7 & 3425535799257924608 & 92.7942 & 24.2934 & -7.17 & 1.09 & \textit{APOGEE}\\
8 & 3425403823502917888 & 92.2634 & 23.8036 & -3.01 & 11.97 & \textit{Gaia DR3}\\ 
9 & 3425404751215869184 & 92.1545 & 23.8343 & -11.77 & 8.84 & \textit{Gaia DR3}\\
\ldots & \ldots & \ldots & \ldots & \ldots & \ldots & \ldots \\
366 & 3426161863753205888 & 91.8309 & 23.9062 & -1.67 & 1.76 & \textit{LAMOST}\\ 
367 & 3426285494391048576 & 92.0642 & 24.4241 & -3.97 & 0.56 & \textit{LAMOST}\\ 
368 & 3426314940686657920 & 92.2238 & 24.7137 & 5.80 & 3.83 & \textit{LAMOST}\\
\hline
\end{tabular}
\end{table*}

The mean radial velocity of NGC~2168 was calculated using a weighted mean approach following the methodology described by \citet{Carrera_2022}. From the full sample of 368 stars, we derived a final mean radial velocity of $-7.60 \pm 0.14$~km~s$^{-1}$. The individual stellar measurements, their associated uncertainties, and survey origins are reported in Table~\ref{tab:rv_sample}, while the adopted mean value is summarized in Table~\ref{tab:center}. As an independent comparison, \citet{Leiner2015} identified 263 probable member stars of NGC~2168 with high-quality radial velocity measurements, yielding a mean value of $-8.16$~km~s$^{-1}$. This value is fully consistent with our determination within the combined uncertainties, supporting the robustness of the derived cluster radial velocity. Using the adopted radial velocity together with the tangential velocity inferred from \textit{Gaia} astrometry, we estimate the total space velocity of the cluster as $v_{\rm space} = \sqrt{v_r^2 + v_t^2} \simeq 16.81$~km~s$^{-1}$.  These kinematic constraints allow us to compute the orbital parameters of the cluster, which are presented in the next subsection (Section~\ref{sec:orbit}).
\begin{table*}
    \centering
    \footnotesize
    \caption{The kinematic parameters of NGC~2168. The parameter values are obtained through Gaussian fitting, as illustrated in Fig.~\ref{Fig:param_fit}.}
    \label{tab:center}
    \setlength\tabcolsep{2.pt} 
    \begin{tabular}{ccccccccc}
\hline \hline 
$\alpha$ & $\delta$ & $\mu_{\alpha}\cos\delta$ & $\mu_{\delta}$ & ${\varpi}$ & Distance & $v_t$ & $\theta$ & $v_r$ \\
        (deg) & (deg) & (mas yr$^{-1}$) & (mas yr$^{-1}$) & (mas) & (pc) & (km s$^{-1}$) & (deg) & (km s$^{-1}$) \\
        \midrule
        92.26 $\pm$ 0.11 & 24.30 $\pm$ 0.10 & 2.278 $\pm$ 0.006 & $-2.893 \pm 0.006$ & 1.154 $\pm$ 0.052 & 840 $\pm$ 54 & 14.70 $\pm$ 0.81 & $-51.85 \pm 2.43$ & $-7.60 \pm 0.14$ \\
\hline
    \end{tabular}
\end{table*}

%

%
%
\subsection{Cluster Orbital Motion in the Milky Way} \label{sec:orbit}

OCs are remarkable markers for understanding the evolution of the Galactic disc. Utilizing $Gaia$ DR3, researchers can analyze their kinematics with exceptional precision, particularly in terms of proper motion and parallax ($\mu_{\alpha} \cos \delta$,~  $\mu_{\delta}$,~ and $\varpi$). 
Furthermore, \textit{Gaia} DR3 offers radial velocities ($V_r$) for millions of relatively luminous, late-type stars \cite{Sartoretti2018}, gathered by the Radial Velocity Spectrometer (RVS) instrument \cite{Cropper2018}. The integration of parallax, proper motion, and RV yields significant phase-space data.  For example, \cite{Antoja2018} demonstrated the great potential of $Gaia$ data for studying the kinematics of the Galactic disc and OCs, revealing the richness of phase-space substructures \citep{Yontan2023, Bisht2026}.

OCs trace the formation and evolution of our Galaxy. Their ages cover the entire lifespan of the Galactic disc, spanning the young to old thin-disc components. Their spatial distribution and motion help to better understand the gravitational potential and the perturbations acting on the structure and dynamics of the Galaxy \citep{canbay2025,Cinar2026}. The orbital motions of OCs are essential not only for understanding their dynamical evolution in the Galaxy but also for investigating the dynamics of the Galaxy itself \citep{Tasdemir2025}. 

To compute a cluster's orbit, we must first adopt a model for the Galactic potential. This potential must accurately reproduce the observed mass density of the Galaxy. For this purpose, we performed the backward orbital integration of NGC 2168 using the \textsc{“MWPotential2014”} model, the default Galactic potential in the \textit{galpy} package \cite{Bovy2015}. This model is made up of three parts:  \textbf{(1)} the bulge component, described by a spherical power-law potential \cite{Bovy2015}, \textbf{(2)} the Galactic disk potential, defined by the Miyamoto-Nagai expression \cite{Miyamoto1975}, and \textbf{(3)} the dark matter halo potential, given by the Navarro-Frenk-White profile \cite{Navarro1995}. The Sun’s galactocentric radius, orbital velocity, and $z$-coordinate were taken as $R_{GC}=8$ kpc, $V_{rot}=220$ km $s^{-1}$, and $z=20.8$ pc \cite{Bovy2015}. \textcolor{black}{In addition, the peculiar velocity of the Sun was corrected using the (U, V, W)$_\odot$ = (11.1, 12.24, 7.25) km s$^{-1}$ values from \citet{Binney2010}, which were applied to account for the Sun’s motion relative to the local standard of rest in our kinematic calculations.}

\begin{figure*}
   \centering
   \includegraphics[width=11.cm, angle=0]{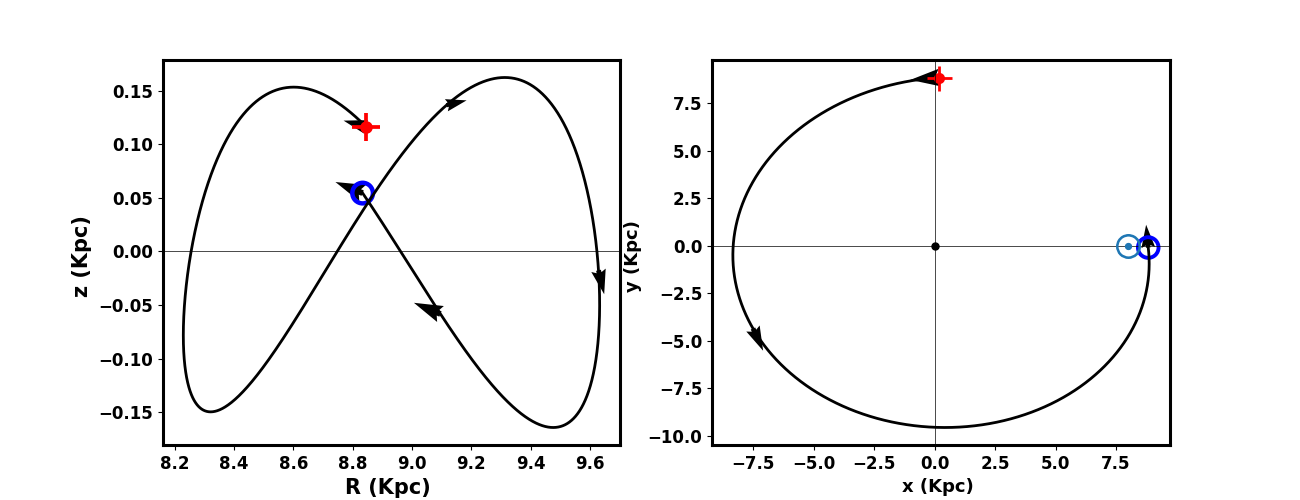}
   \caption{The cluster orbit. The red cross is the birth place. The open blue circle is the current place. The cluster moves in Galactic center direction within Galactic plane, and as a result, it generally faces the impact of Galactic tidal forces.} 
     \label{Fig:orbit}
   \end{figure*}

For input, we used the cluster parameters presented: proper motions ($\mu_{\alpha} \cos \delta$, $\mu_{\delta}$), distance from the Sun, equatorial coordinates ($\alpha$, $\delta$), and radial velocity, which was calculated as an average from the \textit{Gaia DR3} data for member stars. Fig.~\ref{Fig:orbit} shows the integrated orbit of NGC 2168 in the Cartesian Galactocentric coordinate system, backward in time according to the cluster age determined in this study. The red cross indicates the birthplace of the cluster. 

The orbit remains confined close to the Galactic mid-plane, reaching a maximum vertical excursion of $z_{\rm max}=0.171$ kpc. The characteristic orbital period derived from our integration is $T_p=0.253$ Gyr. Therefore, NGC 2168  belongs to the very thin-disk component of the Galaxy. The apocenter $R_{apo}$ and the pericenter $R_{peri}$ are found to be 9.63 and 8.23 Kpc, respectively, which correspond to the eccentricity of the orbit ($e=( R_{apo}-R_{peri})/(R_{apo}+R_{peri})$) 0.08.
The corresponding heliocentric velocity components $(U,V,W)$ and the Galactocentric Cartesian velocity components $(v_x,v_y,v_z)$ derived from the adopted astrometric solution are summarized in Table~\ref{tab:orbit}.

\begin{figure*}
   \centering
   \includegraphics[width=14cm, angle=0]{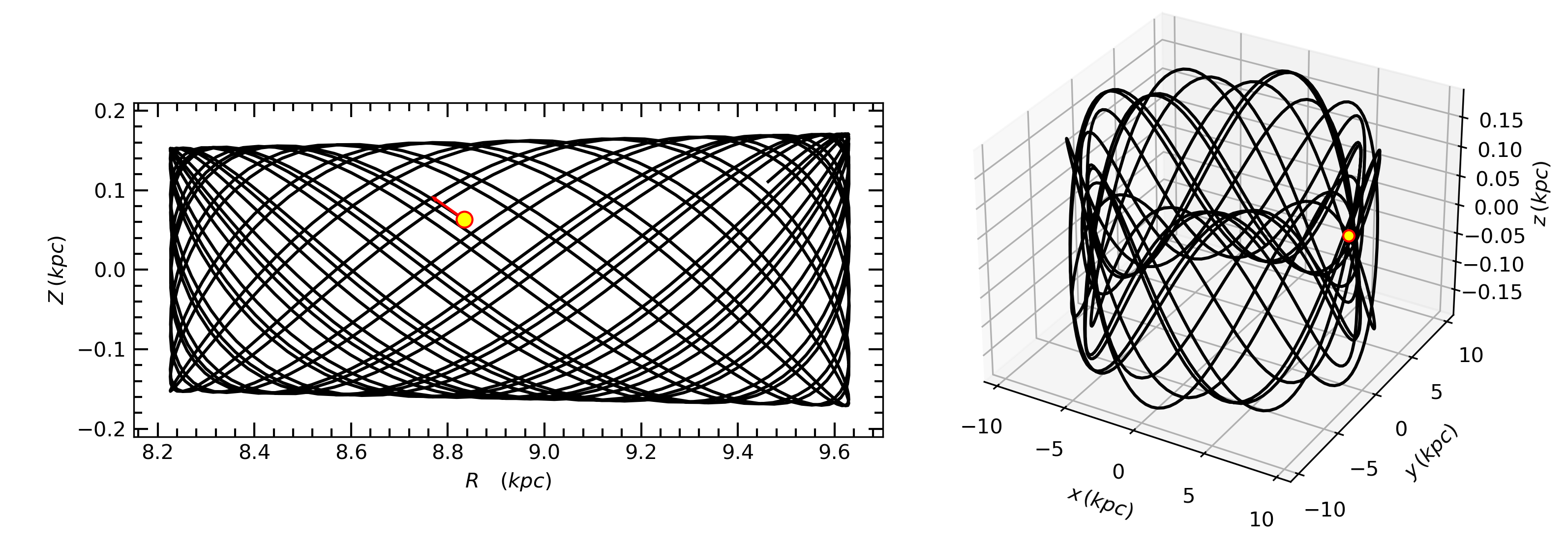}
   \caption{Galactic orbit of the cluster integrated for 3 Gyr in the adopted Milky Way potential to examine the long-term, quasi-periodic orbital behaviour and the confinement of the trajectory in configuration space. The left panel displays the motion in the meridional plane ($R_{\rm GC}$ x $Z$), illustrating the vertical excursion as a function of Galactocentric radius. The right panel shows the three-dimensional trajectory in Galactocentric Cartesian coordinates ($X,Y,Z$). The filled yellow marker indicates the present-day position.} 
     \label{Fig:orbit3gyr}
   \end{figure*}

In addition, Fig.~\ref{Fig:orbit3gyr} shows the Galactic orbit of NGC 2168 integrated for 3 Gyr in the adopted Milky Way potential. This extended integration is used to illustrate the long-term orbital morphology and to verify that the trajectory exhibits recurrent (quasi-periodic) behaviour in the adopted potential. The meridional-plane projection ($R_{\rm GC}\times Z$) indicates that the cluster remains confined to low $|Z|$ while undergoing modest radial excursions from $R_{\rm peri}$ to $R_{\rm apo}$, and the 3D trajectory further highlights the low-eccentricity nature of the orbit over long timescales. The birthplace is defined separately at $t=-\tau$ (Fig.~\ref{Fig:orbit}), whereas Fig.~\ref{Fig:orbit3gyr} is shown for long-term dynamical context.

%
%

To summarize this section, the NGC 2168 cluster is located 8.84 kpc from the Galactic center and is moving parallel to the Galactic plane towards the Galactic center, following a nearly circular orbit.

\begin{table*}[t]
\centering
\caption{Derived Galactic orbital parameters of NGC~2168.}
\label{tab:orbit}
\begin{tabular}{ccccc|cccccc}
\hline\hline
$R_{\rm apo}$ & $R_{\rm peri}$ & $e$ & $R_{\rm gal}$ & $T_p$ &
$v_x$ & $v_y$ & $v_z$ & $U$ & $V$ & $W$ \\
(kpc) & (kpc) &  & (kpc) & (Gyr) &
\multicolumn{6}{c}{(km~s$^{-1}$)} \\
\hline
9.63 & 8.23 & 0.08 & 8.84 & 0.253 &
$-20.41$ & $218.76$ & 9.29 & 9.31 & $-13.47$ & 2.08 \\
\hline
\end{tabular}
\end{table*}

%
%
\subsection{The Elongation of NGC 2168}

The evolution of an open cluster, influenced by internal or external forces, is reflected in its changes in shape. More than a century ago, the flattening of a moving cluster was postulated by \cite{Jeans1916}. An important aspect of OCs is their morphological structure, which is associated with features such as elongated shapes and tidal tails. The orientation of cluster elongation or a tidal tail reflects the direction and nature of the gravitational tidal force.

\begin{figure*}[t]
\centering
\subfloat[The co-moving stars of members which are with $r_{cl}$. The NGC 2168 cluster exhibits a distinctly elongated structure oriented not along the direction of orbital motion, however rather perpendicular to the Galactic plane, as depicted in this image.]{\label{fig:comov_elong}\includegraphics[width=.3\linewidth]{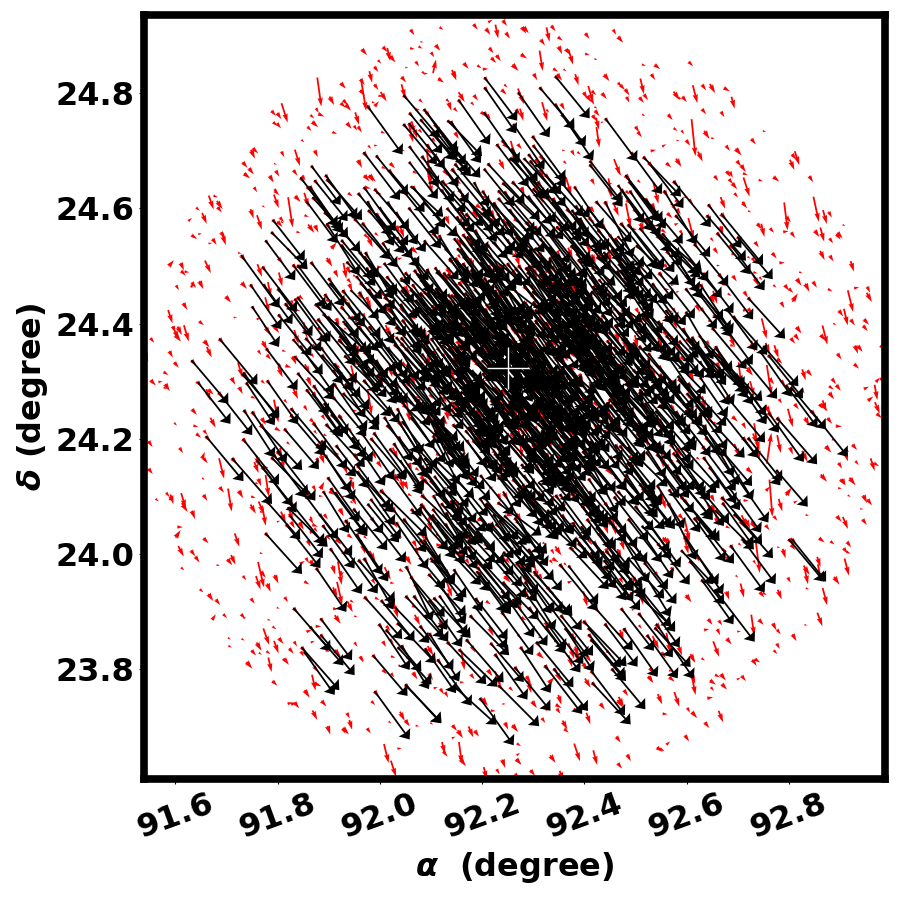}}\hfill
\subfloat[The co-moving stars of members from \cite{Kos2024}, illustrate the nuclei and halos of widely spread stars.]{\label{fig:comov_Kos}\includegraphics[width=.3\linewidth]{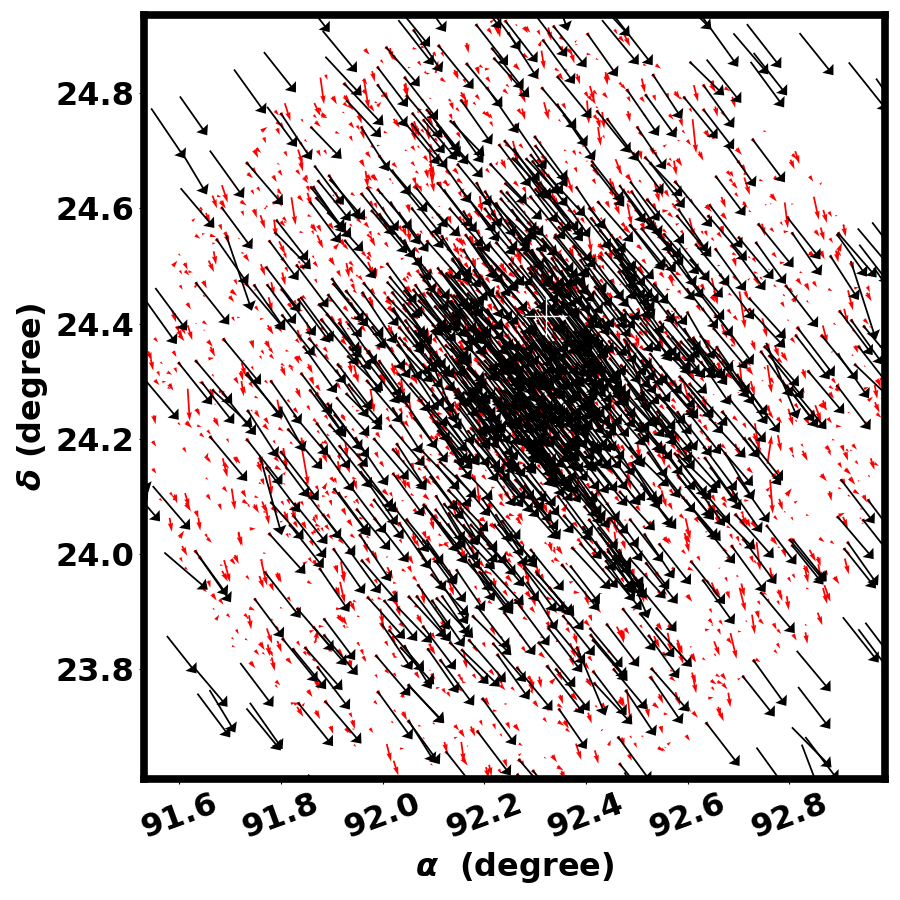}}\hfill
\subfloat[The co-moving stars belonging to members cited in \cite{Cantat-Gaudin2020} are used to represent the elongation of the NGC 2168 cluster \cite{Hu2021a}, however the elongation is not particularly clear.]{\label{fig:comov_Cantat}\includegraphics[width=.3\linewidth]{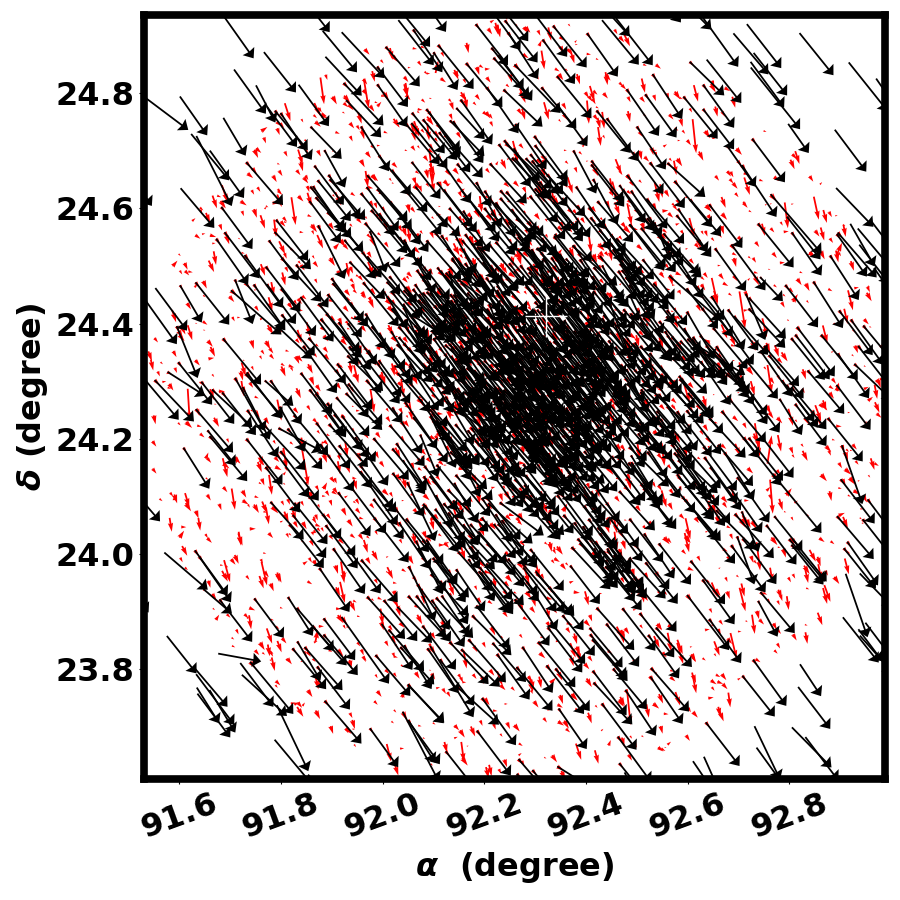}}\hfill
\caption{The co-moving star diagrams of NGC 2168, derived from $Gaia$ DR3, are based on members from our research as well as other sources.}
\label{Fig:comov}
\end{figure*}

The analysis conducted by \cite{Kos2024} on 476 OCs utilized a probabilistic approach, revealing elongated structures in all samples, with NGC 2168 being one of them. The co-moving stars associated with the members identified by \citet{Kos2024} are illustrated in Fig.~\ref{fig:comov_Kos}. In addition, the RDP of these members is illustrated in Fig.~\ref{Fig:test_king}. 

\cite{Hu2021a} conducted an analysis of the morphology of 1256 OCs employing nonparametric bivariate density estimation. They found elongated morphologies in the majority of their samples, with NGC 2168 being one of them. The co-moving star diagram for their members is depicted in Fig.~\ref{fig:comov_Cantat}.

In this research, we also find an elongated structure, as depicted in Fig.~\ref{fig:comov_elong}, using a method that combines the HDBSCAN membership probability with the King-model-based structural characterization. Furthermore, we observe that the cluster's elongation is oriented almost perpendicular to the Galactic plane, contrasting with the typical alignment along the orbital path expected from differential rotation \citep{Kos2024}. This vertical deformation likely arises from the cluster's interaction with the Galactic disk potential. As our orbital analysis indicates (Section~\ref{sec:orbit}), NGC 2168 remains confined close to the Galactic mid-plane with a maximum vertical excursion of $z_{\max}\approx171$~pc and follows a nearly circular orbit (Table~\ref{tab:orbit}). Consequently, the observed perpendicular elongation may be a signature of vertical tidal heating or disk shocking associated with repeated passages through the Galactic plane, which can inject energy into the system and imprint vertical distortions on the stellar distribution.


\section{Summary and Conclusions} \label{sec:summary}

In this work, we revisited the fundamental properties, structural characteristics, and dynamical state of the open cluster NGC~2168 (M35) by exploiting the astrometric precision and photometric depth provided by \textit{Gaia}~DR3. A persistent obstacle in earlier analyses has been the substantial contamination introduced by the nearby and denser background cluster NGC~2158. By capitalizing on the clear parallax separation between the two systems ($\Delta \varpi \sim 0.9$ mas), we were able to isolate the NGC~2168 population with high confidence, thereby establishing a robust membership catalog that is essential for deriving reliable cluster parameters.

The isochrone solutions obtained in this study yield an age of $190 \pm 12$ Myr and a distance of $850 \pm 62$ pc. Our age estimate is slightly older than the $\sim$150 Myr values commonly quoted in earlier ground-based photometric studies \citep{Sung1999, Kalirai2003}, yet it falls squarely within the range reported by recent \textit{Gaia}-driven analyses such as \cite{Cantat-Gaudin2020a} and \cite{hunt2024improving}, which find 178 Myr and 166 Myr, respectively (see Table~1). The discrepancy with older work likely reflects the improved removal of field interlopers and NGC~2158 contaminants, which had previously broadened the cluster’s main-sequence turn-off region and biased age determinations. The metallicity we obtain, $[M/H] \approx -0.048$ dex, agrees closely with spectroscopic measurements and reinforces the view that NGC~2168 is chemically near-solar.

Our structural analysis shows that NGC~2168 is markedly more extended than reported in pre-\textit{Gaia} studies. The generalized King model with $\beta = 1$ yields a limiting radius of $r_{cl} \approx 36.7$ arcmin, significantly exceeding the $\sim 15'$ reported by \cite{Bouy2015}. This difference is best interpreted as the recovery of the cluster’s diffuse halo, which earlier surveys could not reliably detect against the background. The shallower fall-off implied by the $\beta = 1$ profile further suggests the presence of an extended outer envelope, a signature often associated with ongoing dynamical evolution, mass loss, and interaction with the Galactic tidal field.

The orbital parameters we derive ($e = 0.08$ and $z_{\max} \sim 171$ pc) place NGC~2168 firmly within the thin-disk population. Yet, the spatial morphology displayed in Fig.~\ref{Fig:comov} (Panel A) reveals a pronounced elongation perpendicular to the Galactic plane. Because tidal tails of open clusters generally align with their orbital trajectories \citep{Kos2024}, this orthogonal deformation may point to vertical disk heating, external perturbations, or a recent dynamical disturbance. The observed structure, therefore, indicates that NGC~2168 is dynamically active and continues to respond to the larger-scale Galactic potential.

Using \textit{Gaia}~DR3 astrometry supplemented by 2MASS photometry, we conducted a detailed chemo-dynamical and structural examination of the intermediate-age open cluster NGC~2168. Our main results are as follows:

\begin{itemize}
\item A strict parallax cut ($0.9 \le \varpi \le 1.4$ mas) efficiently removed contamination from NGC~2158 and the field. Spatial density diagnostics confirm that the resulting sample represents a clean and reliable cluster sequence.

\item PARSEC isochrone fitting to the decontaminated CMD yields an age of $190 \pm 12$\,Myr, a metallicity of $[M/H] \approx -0.048$\,dex, and a distance of $850 \pm 62$\,pc. These values are fully consistent with the independent distance inferred from parallax inversion ($840 \pm 54$\,pc).

\item The radial density profile is best reproduced by a generalized King model, with a core radius of $r_c = 7.97 \pm 1.20$\, arcmin and a limiting radius of $r_{cl} = 36.69 \pm 1.86$\, arcmin. Both the mass function and the extended density profile indicate that the cluster retains a substantial stellar population in its outer halo. Overall, the relatively high concentration parameter ($C = 4.60$) and density contrast ($\delta_c = 8.45$) indicate that NGC~2168 is a dynamically evolved open cluster whose present-day structure reflects significant internal relaxation and mass segregation processes, placing it among the more centrally concentrated intermediate-age Galactic open clusters when compared to similar systems in the literature \citep{King1966, Bonatto2007}.

\item The mean proper motion is $(\mu_{\alpha}\cos\delta, \mu_{\delta}) = (2.278, -2.893)$\,mas\,yr$^{-1}$. Orbital integration shows that NGC~2168 follows a nearly circular thin-disk orbit ($e = 0.08$) and experiences vertical oscillations with a period of $\sim 9.88$\, Myr.

\item  The spatial distribution of co-moving stars reveals a mild but coherent elongation. This structure may reflect the cluster’s dynamical interaction with the Galactic disk and suggests that deeper surveys are needed to trace potential tidal features beyond the current detection limits.

\end{itemize}

\section*{Acknowledgments}
We thank the two anonymous referees for their careful reading of the manuscript and for their constructive comments, which have improved the clarity and quality of this work. This work has made use of data from the European Space Agency (ESA) space mission Gaia. Gaia data are being processed by the Gaia Data Processing and Analysis Consortium (DPAC). Funding for the DPAC is provided by national institutions, in particular the institutions participating in the Gaia MultiLateral Agreement (MLA). The Gaia mission website is \textcolor{black}{https://www.cosmos.esa.int/gaia}. The Gaia archive website is \textcolor{black}{https://archives.esac.esa.int/gaia}. 

We gratefully acknowledge the Python community for developing the open-source software tools used in this research, specifically for Matplotlib, Numpy, Scipy and Astropy etc. Their efforts have contributed to making data analysis easier as well as representing it graphically in a creative way.

This publication makes use of data products from the Two Micron All Sky Survey, which is a joint project of the University of Massachusetts and the Infrared Processing and Analysis Center California Institute of Technology, funded by the National Aeronautics and Space Administration and the National Science Foundation.


\bibliographystyle{jasr-model5-names}
\biboptions{authoryear}
\bibliography{Reference}

\end{document}